\documentclass{aa}  
\usepackage{graphicx}
\usepackage{multirow}
\usepackage{booktabs}
\usepackage{xcolor}
\usepackage{natbib,twoopt}
\usepackage[breaklinks=true]{hyperref} 
\bibpunct{(}{)}{;}{a}{}{,}             
\makeatletter
  \newcommandtwoopt{\citeads}[3][][]{\href{http://adsabs.harvard.edu/abs/#3}%
    {\def\hyper@linkstart##1##2{}%
     \let\hyper@linkend\@empty\citealp[#1][#2]{#3}}}
  \newcommandtwoopt{\citepads}[3][][]{\href{http://adsabs.harvard.edu/abs/#3}%
    {\def\hyper@linkstart##1##2{}%
     \let\hyper@linkend\@empty\citep[#1][#2]{#3}}}
  \newcommandtwoopt{\citetads}[3][][]{\href{http://adsabs.harvard.edu/abs/#3}%
    {\def\hyper@linkstart##1##2{}%
     \let\hyper@linkend\@empty\citet[#1][#2]{#3}}}
  \newcommandtwoopt{\citeyearads}[3][][]%
    {\href{http://adsabs.harvard.edu/abs/#3}
    {\def\hyper@linkstart##1##2{}%
     \let\hyper@linkend\@empty\citeyear[#1][#2]{#3}}}
\makeatother
\usepackage{txfonts}

\newcommand{\Ko}{$\mathrm{K^0}$}
\newcommand{\Ks}{$\mathrm{K_S}$}

\begin{document}

   \title{Extinction law and stellar mass in the Nuclear Bulge from kinematically-selected red clump stars\thanks{The full version of Table \ref{tab:redd_map} is only available in electronic form at the CDS via anonymous ftp to cdsarc.cds.unistra.fr () or via https://cdsarc.cds.unistra.fr/viz-bin/cat/J/A+A/VOL/PAGE.}}

   \subtitle{}

   \author{Á.~Valenzuela~Navarro\inst{1} \fnmsep
            \thanks{email: \href{mailto:auvalenzuela@uc.cl}{auvalenzuela@uc.cl}}
          \and
          M.~Zoccali \inst{1}
          \and
          E.~Valenti \inst{2,3}
          \and
          R.~Contreras~Ramos \inst{1}
          \and
          A.~Rojas-Arriagada\inst{4,5}
          \and
          A.~Luna \inst{6}
          \and
          R.~Albarracín \inst{7,8}
          \and
          C.~Gallart \inst{9,10}
          \and
          J.~Olivares~Carvajal \inst{11}
          \and
          F.~Gran \inst{12,13,1}
          \and
          C.~Salvo-Guajardo \inst{1}
          \and
          G.~Nandakumar \inst{13,14}
          \and
          A.~Renzini \inst{15}
          }

   \institute{
             Instituto de Astrofísica, Pontificia Universidad Católica de Chile, Av. Vicuña Mackenna 4860, Santiago, Chile 
         \and 
             European Southern Observatory, Karl Schwarzschild-Straße 2, D-85748 Garching bei München, Germany 
         \and
             Excellence Cluster ORIGINS, Boltzmann-Straße 2, 85748 Garching bei München, Germany 
         \and
             Departamento de Física, Universidad de Santiago de Chile, Av. Victor Jara 3659, Santiago, Chile 
         \and
             Center for Interdisciplinary Research in Astrophysics and Space Exploration (CIRAS), Universidad de Santiago de Chile, Santiago, Chile 
         \and
              European Southern Observatory, Alonso de Córdova 3107, Vitacura, Casilla 19001, Santiago, Chile 
         \and
             Max Planck Institute for Astronomy, D-69117 Heidelberg, Germany 
         \and
             Fakultät für Physik und Astronomie, Universität Heidelberg, Im Neuenheimer Feld 226, 69120 Heidelberg, Germany 
         \and
             Instituto de Astrofísica de Canarias, La Laguna, Tenerife, Spain 
         \and
             Departamento de Astrofísica, Universidad de La Laguna, Tenerife, Spain 
         \and
             INAF - Osservatorio Astronomico di Capodimonte, salita Moiariello 16, 80131, Naples, Italy 
         \and
             Université Côte d’Azur, Observatoire de la Côte d’Azur, CNRS, Laboratoire Lagrange, Blvd de l’Observatoire, 06304, Nice, France 
         \and 
            International Gemini Observatory/NSF NOIRLab, Casilla 603, La Serena, Chile 
         \and
             Aryabhatta Research Institute of Observational Sciences, Manora Peak, Nainital 263002, India 
         \and
             Division of Astrophysics, Department of Physics, Lund University, Box 118, SE-22100 Lund, Sweden 
         \and
             INAF – Osservatorio Astronomico di Padova, Vicolo dell’Osservatorio 5, I-35122 Padova, Italy 
            }
            
   \date{Received November 25, 2025; accepted April 29, 2026}

  \abstract
   {The Nuclear Bulge of the Milky Way harbors stellar populations that provide crucial insights into galaxy formation processes and serve as a nearby analog for understanding bulge formation in external galaxies. However, detailed studies of this region are severely hampered by extreme and highly variable interstellar extinction, which obscures the intrinsic stellar properties and impedes accurate stellar mass determinations.}
   {Our goal is to measure the extinction law towards the Nuclear Bulge and to estimate its stellar density.}
   {We developed a method to determine the extinction law towards the Nuclear Bulge by kinematically selecting red clump stars belonging to this region. We created a high-spatial resolution reddening map, and computed stellar mass with completeness-corrected red clump star counts, scaled from empirical measurements.}
   {We find a total-to-selective extinction ratio of $\mathrm{A_K/{E_{H-K}} = 1.259 \pm 0.074}$, and an extinction ratio of $\mathrm{A_H/A_K = 1.794 \pm 0.046}$, which are consistent with previous works. The high-spatial resolution reddening map shows clear filamentary structures, and a gradient in the extinction over the giant molecular cloud G0.253+0.016 (i.e., the Brick). From the star counts, we measured a stellar mass of $\mathrm{12.2~\pm2.6\times10^8~M_{\odot}}$ for the Nuclear Bulge, in agreement with other mass estimates.}
   {}

   \keywords{Galaxy: center --
             Stars: Hertzsprung-Russell and C-M diagrams --
             Dust, extinction --
             Infrared: stars
            }

   \maketitle
   \nolinenumbers

\section{Introduction}\label{sec:intro}

The MW is a barred spiral galaxy, with the Sun located within the disk at $\sim 8.2$ kpc from its center \citep{gravitycollab+2019, gravitycollab+2022}, and its region within $\sim 3$~kpc from the center is known as the bulge. It has a stellar mass of $\sim \mathrm{~2\times10^{10}~M_{\odot}}$ \citep[e.g.][]{portail+2015, valenti+2016, simion+2017}, which represents about 1/3 of the total stellar mass of the Galaxy. The MW bulge is composed mainly of stars over 10 Gyr \citep[e.g.][]{zoccali+2003, renzini+2018}, meaning that it is the first massive structure to have formed in the MW. Therefore, it holds vital information about its mass assembly history that we must understand to know how and when the bulge formed \cite[see][for a review]{zoccali&valenti2024}.

There is still debate over whether all the stars in the MW bulge were formed \textit{in situ} or if some fraction of them resulted from the accretion of external stellar systems. Notably, within the bulge lies a small ($\sim$200~pc radius) star-forming region near the Galactic Center (GC), called the “Nuclear Bulge” \citep[NB,][]{mezger+1996}. This region is formed of gas-phase structures and stars; the former is known as the Central Molecular Zone (CMZ), and the latter contains several stellar substructures, such as young massive star clusters (i.e., Arches and Quintuplet), a Nuclear Star Cluster (NSC), and an extended component usually referred to as the Nuclear Stellar Disk \citep[NSD; all references bellow, from][]{schoenrich+2015}. Hereafter, we will use the term NB to refer to the entire region, although we will focus on the stellar content. We prefer to avoid using the term NSD, as the detailed morphology and kinematic structure of this region remain under active investigation.  Evidence exists for additional components, including an inner bar \citep{alard2001, rodriguez-fernandez&combes2008}, and questions persist regarding the kinematic properties and whether a distinct disk-like component can be unambiguously identified \citep[see Sect. 6 in][]{zoccali+2024}. For a comprehensive review of the Galactic NSD and the ones in external galaxies, see \citet{schultheis+2025}.

Recent simulations indicate that these nuclear structures form through secular evolution after the bulge has formed a bar. The bar changes the gravitational potential, which funnels gas along the bar's main axis. Once the gas reaches the center, star formation is triggered, forming nuclear discs, rings, or clusters \citep[see e.g.][]{fragkoudi+2016, sormani+2020a, tress+2020}. Indeed, some of the stars within the NB could have been formed from the gas from the CMZ \citep{sormani+2020a, schultheis+2021}. Therefore, if we can measure the ages of the first stars that compose the NB, we can estimate an epoch for the formation of the bar of the Milky Way \citep[e.g.][]{baba+kawata2020, Bittner+2020, de-sa-freitas+2022, de-sa-freitas+2025}. In turn, dating the MW bar is important as it constrains galactic formation models and cosmological simulations.

The nuclear structures observed in the MW have also been detected in other galaxies \citep[e.g.,][]{gadotti+2019, schultheis+2025}, and the NB can help in understanding star formation in distant galaxies. The physical conditions of the gas structures within the CMZ resemble those of high-redshift galaxies in terms of temperature, density, velocity dispersion, and magnetic fields. Unlike those distant galaxies, where individual structures cannot be resolved, the CMZ lies in our own cosmic backyard, allowing us to study star formation processes in unprecedented detail under these physical conditions. For this reason, \citet{kruijssen+longmore2013} referred to the CMZ as a “high redshift analog” star-forming region. 

Furthermore, the CMZ presents a unique opportunity to understand scaling relations. One of these relations is known as the Schmidt-Kennicutt (SK) law \citep{schmidt1959, kennicutt1998}, which correlates the star formation rate (SFR) surface density with the total gas surface density. Although this relation holds for many physical scales, i.e., from nearby star-forming regions to distant starburst galaxies, \citet{longmore+2013a} found that the CMZ lies one order of magnitude below the expected value of SFR density given the SK law. This could indicate that star formation depends on other environmental factors rather than the cloud density \citep[see e.g.][and references therein]{henshaw+2023}. Solving this puzzle could potentially enhance our understanding of the details of star formation processes and their interactions with the environment.

Given the relevance of the NB for all these different aspects of astrophysics, it is crucial to study the stellar populations that compose it in depth. However, any attempt to observe the NB must have high spatial resolution, given the high crowding levels towards this particular direction in the sky. Unfortunately, high spatial resolution is often coupled with a small field of view, which is why deep space-based observations are available only for some sparse fields \citep[e.g.][]{hosek+2022, schoedel+2023}. Given that this region has a high surface density of stars and extreme extinction, optical wavelengths are virtually blind to the stellar populations in the NB. To solve this, we must use high-spatial resolution near-infrared (NIR) photometry, often coupled with adaptive optics \citep[see e.g.][analyzing the Arches and Quintuplet clusters]{stolte+2015}. All these difficulties have deterred comprehensive and homogeneous observational campaigns in the NB region.

Nevertheless, in recent years, the GALACTICNUCLEUS project \cite[GNC,][]{nogueras-lara+2019b} overcame the aforementioned observational difficulties and studied the NB region in detail. For the first time, a dedicated observational campaign produced deep NIR catalogs for J, H, and \Ks\ filters, extinction ratios, reddening map, and star formation history (SFH) estimations for stars in the NB, over an area of 0.3 deg$^2$ \citeauthor{nogueras-lara+2019b}(\citeyear{nogueras-lara+2019b, nogueras-lara+2020a, nogueras-lara+2020b, nogueras-lara+2021c}). They achieved all this by using HAWK-I NIR imager \citep{pirad+2004, casali+2006, kissler-patig+2008, siebenmorgen+2011} located at the Very Large Telescope (VLT) UT-4 telescope in Cerro Paranal, using the speckle holography photometry technique described in \citet[][]{schoedel+2013}.

\begin{figure*}[ht]
 \centering
 \includegraphics[width=0.85\textwidth]{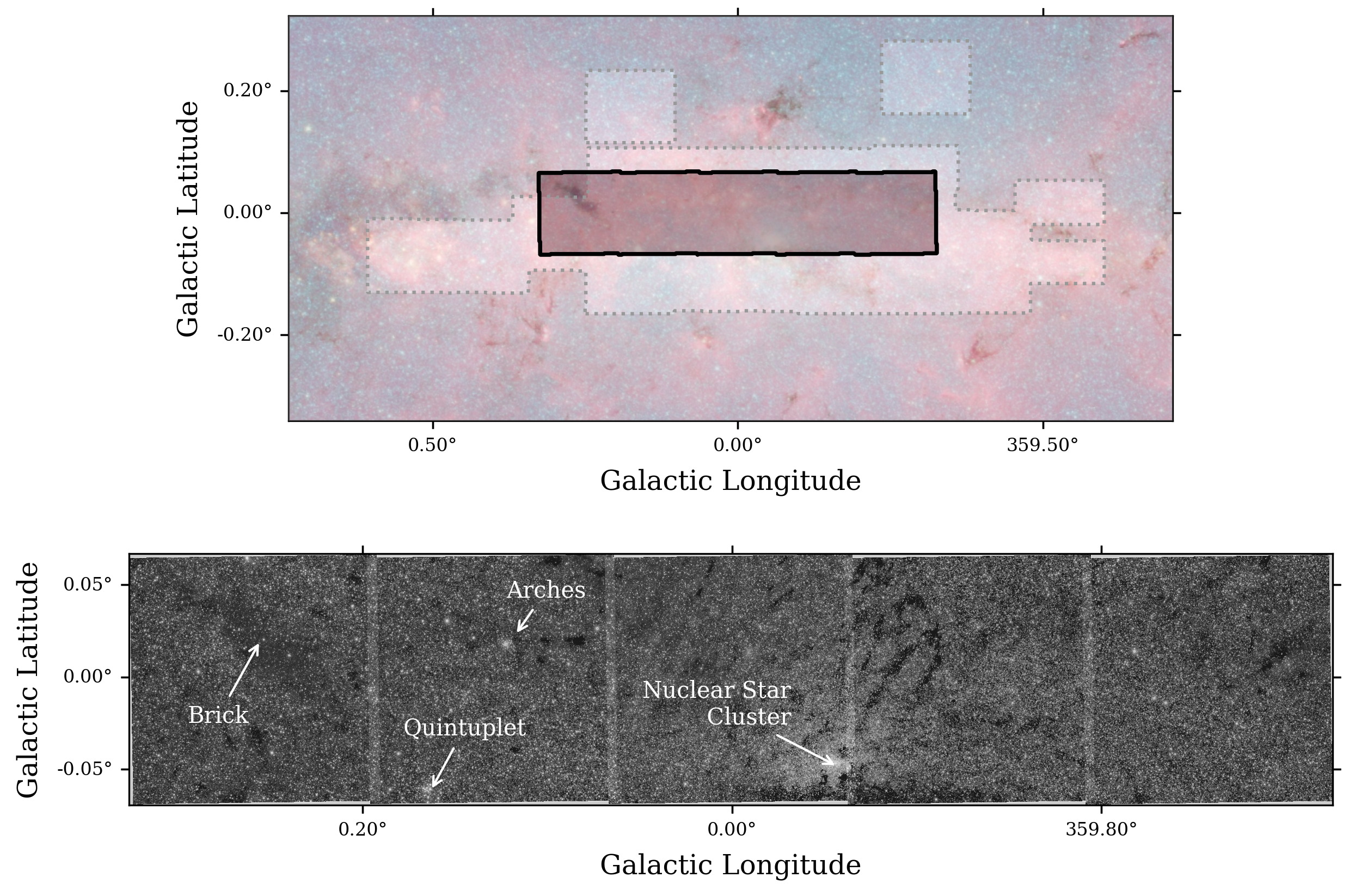}
 \caption{\textit{Top:} Image of the NB of the MW from GLIMPSE360 survey \citep{whitney+2011} from Spitzer Space Telescope. The red color represents 4.5 $\mu m$, and the blue color 3.6 $\mu m$. The black line surrounding the dark-shaded area indicates the footprint of our dataset, as shown in the bottom panel. The gray dotted line encompassing the white-shaded area depicts the GNC survey area. \textit{Bottom:} Five HAWK-I pointings in \Ks\ filter, covering the innermost part of the NB. Some relevant structures are indicated with arrows. The white vertical lines show the overlapping regions between the adjoint HAWK-I pointings.}
 \label{fig:nb_image}
\end{figure*}

Overlapping with the GNC survey, a similar archival dataset was obtained using the HAWK-I imager (ID 0103.B–0262(A), PI: Zoccali, hereafter referred to as NB Field, NBF). Unlike the observation strategy adopted by GNC, we did not use speckle holography, and obtained shallow and deep observations in five fields\footnote{Only five fields were observed of the 124 originally requested} centered at $b = 0^{\circ}$ between $|l| < 0.37^{\circ}$, covering approximately 320 arcmin$^2$. Both datasets provide an opportunity for independent verification of the scientific results in the NB. The spatial coverage of both datasets is shown in the top panel of Fig.~\ref{fig:nb_image}.

One key aspect critical for all the science in the NB is the characterization of the extinction and reddening towards this particular direction of the MW. Previous studies, including GNC survey, have shown that extinction values are severe, which get as high as $A_V \ge 30\ \text{mag}$ and $A_{K_s} \ge 2.5\ \text{mag}$ \citep[e.g.][]{nishiyama+2006, schoedel+2010, nogueras-lara+2020a}. Moreover, there is evidence for a wavelength dependence in the extinction law towards the NB \citep{nogueras-lara+2020a}, which makes it even more critical to a deep understanding of the impact of the dust on stellar observations. Current reddening maps available in this area are done using the red clump (RC) method \citep{nishiyama+2006}, which uses this feature in the observed color-magnitude diagrams (CMDs) to trace stellar structures composed of metal-rich core helium-burning stars \citep[see][for comprehensive review]{girardi2016}. In fact, RC stars have been used to trace galactic structures, such as the galactic bar \citep[e.g.][]{stanek+1994, stanek+1997, wegg&ortwin2013, simion+2017, gonzalez+2018} and also to study the galactic extinction law \citep[e.g.][]{nishiyama+2006, nishiyama+2009, alonso-garcia+2015, alonso-garcia+2017, sanders+2022}. Even more, accurate RC star counts can be used to compute the stellar mass content: \citet{valenti+2016} measured the stellar content of the MW Bulge by using RC stars with an empirical approach, and \citet{simion+2017} did the same assuming a Chabrier initial mass function (IMF). We discuss the latter method in Sect. \ref{sec:stellar_mass}.

In this paper, we use the RC stars as both extinction and structure tracers by assuming that the slope in the observed CMDs does not represent the reddening law, as pointed out by e.g. \citet[][]{nishiyama+2006, nogueras-lara+2019b}, but is the combination of both extinction and distance spread along the line-of-sight (LOS), with different extinction at each layer. To assess this, we used kinematics of RC stars in the NB to derive a new extinction law. This method relies on the fact that stars on the near side of the NSD should be rotating in a direction opposite to that of the far side. From there, we constructed a high-spatial resolution reddening map. Finally, we used completeness-corrected RC star counts to estimate the stellar content of the NB. 

\section{Data}\label{sec:data}

\subsection{Image reduction and processing} \label{sec:image_proc}

We analyzed H and \Ks\ observations obtained with HAWK-I NIR imager at VLT-UT4 (Yepun Telescope) at ESO Paranal Observatory. HAWK-I has a field of view (FoV) of $7.5 \times 7.5$ arcmin, with a spatial resolution of 0.106 arcsec/pix. The dataset comprises five fields observed within program ID 0103.B–0262(A), between July and September 2019. The seeing during the observations was 0.7 arcsec on average, but the GRound layer Adaptive optics system Assisted by Laser (GRAAL) system \citep{arsenault+2008, paufique+2010} improved the Point Spread Function Full Width at Half Maximum (PSF FWHM) to 0.4 arcsec for the H band and 0.3 arcsec for \Ks. The dithering pattern of the observations was designed to homogeneously cover the 15 arcsec gap between the four detectors and to improve sky subtraction.

Processing of the raw data was carried out using the HAWK-I pipeline within the Reflex workflow\footnote{\url{https://ftp.eso.org/pub/dfs/pipelines/instruments/hawki/}}. To account for instrumental signatures, the pipeline first generates a set of calibration products, including a combined dark frame (with the associated bad-pixel map), a master flat field to correct for pixel-to-pixel sensitivity variation, and estimates of the read-noise, gain, and covariance for each HAWK-I detector. These calibrations are applied to each dithered science exposure, after which the sky background – derived from a median combination of the dithered images – is subtracted. Finally, for each filter, the pipeline realigns the corrected science frames and combines them into a stacked image. More details on the pipeline procedure and used algorithms can be found on the ESO pipeline web page\footnote{\url{https://www.eso.org/sci/software/pipe_aem_main.html}}.

The observing strategy consists of long and short exposure images to detect fainter stars in the former, and recover brighter stars in the latter. For the long \Ks\ filter, one stacked image was produced from eight individual images of 10~s exposures with two sub-integrations, giving a total exposure time of 160~s. Whereas for the long H-band, the stacked image was produced from 16 individual 10~s exposures with three sub-integrations, for a total exposure time of 480~s. On the other hand, all the short exposure images for H and \Ks\ filters were produced from five individual images, 2s long with one sub-integration, resulting in a total exposure time of 10s. Field NBF058 has one extra stacked image of 180~s long, produced from six individual 10s images with three sub-integrations. Table~\ref{tab:data_descript} summarizes the data used in this work.

The complete mosaic of all five fields for \Ks\ images can be seen in the bottom panel of Fig.~\ref{fig:nb_image}. Some relevant structures can be spotted just by visual inspection, such as the Nuclear Star Cluster (central tile), and two young massive clusters: Arches and Quintuplet (second tile from the left), and the giant molecular cloud G0.253+0.016 (from now on, the Brick, in the first tile on the left). The former belongs to the tile NBF054, which was presented and discussed in \citet{zoccali+2021}.

\subsection{PSF photometry}\label{sec:photometry}

We performed PSF profile fitting using \texttt{DAOPHOT} \citep{stetson1987}, \texttt{ALLSTAR}, and \texttt{ALLFRAME} software \citep{stetson1994}, on a total of 28 H and \Ks\ images (N$_{\rm obs}$ in Table~\ref{tab:data_descript})\footnote{A Python wrapper is available in GitHub if requested to the corresponding author.}. 
We worked on each of the four HAWK-I detectors (i.e., chips) individually. As a first step, we performed {\it Find} and aperture {\it Photometry} on each chip, with standard parameters (e.g., detector gain, saturation, FWHM). We then drew 200 isolated stars for each chip, which we visually inspected to select the best 30-50 to create a first PSF model. As a second iteration, we subtracted all the neighbors from those stars and created a new, cleaner model, which was used to perform the \texttt{ALLSTAR} photometry. 
The output catalogs were registered using \texttt{DAOMATCH} routine, and stacked using \texttt{MONTAGE2}. A new complete PSF photometry was performed on this stacked image, using the PSF model from a long exposure H band. This catalog was used as a master star list to give as input to \texttt{ALLFRAME}, which produced the final photometry analyzed here. 

\subsection{Photometric calibration}

Our goal is to calibrate the HAWK-I photometry to the 2MASS system \citep{skrutskie+2006}. However, the stars in common between NBFs and 2MASS are either saturated in HAWK-I data or very faint in 2MASS. Also, the difference in spatial resolution makes direct cross-match difficult. Therefore, we used the Vista Variable in the Via Lactea Survey \citep[VVV, ][]{minniti+2010} as an intermediary between the NBF and 2MASS.

First, we calibrated the VVV/VVVX PSF photometry (Contraras Ramos, private communication) to 2MASS following a color-dependent transformation. The color terms (CTs) between 2MASS and VVV are, in fact, non-negligible but small: $\mathrm{CT_H} = 0.048 \pm 0.004$ and  $\mathrm{CT_{K_s}}~=~0.026 \pm 0.007$, in the H and \Ks\ bands, respectively. The zero points (ZP) were found by cross-matching the stars of 2MASS and VVV in the overlapping region with the NBFs, and we report them in Table~\ref{tab:ZP_2MASS_VVV}. As a second step, we used the calibrated VVV catalog as a reference to calibrate the NBF photometry using a color-dependent photometric transformation. The CT and a ZP were applied to each chip if the obtained values were not consistent with zero within the uncertainties. We provide the technical details of the calibration in Appendix~\ref{sec:appendix_phot}.

After calibrating each individual NBF chip, we merged the 2MASS-calibrated NBF photometry for individual HAWK-I chips into a single catalog, with 974\,361 stars. In the overlapping regions of NBFs, the quoted magnitude is the mean of the measured magnitude on each chip, and the errors are added in quadrature. Hereafter, the calibrated NBFs \Ks\ magnitude will be labeled as K for simplicity. 

\begin{figure}[]
 \centering
 \includegraphics[width=0.90\hsize]{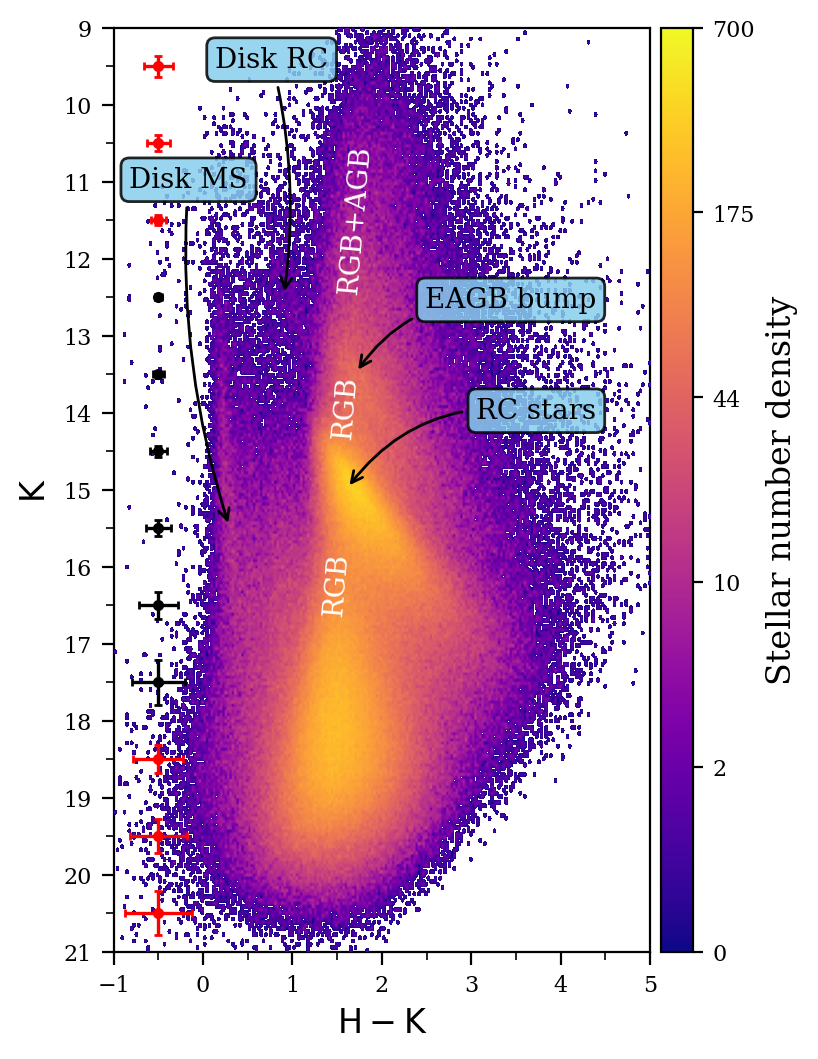}
 \caption{CMD of all the NBFs colored by density. Key observed features are labeled. The colorbar shows the stellar number density, and the color scale is on a logarithmic scale to enhance the visibility of other features less populated than the RC. The typical photometric errors are shown on the left. The red dots are the photometric errors reported by \texttt{DAOPHOT/ALLFRAME}, whereas the black error bars are computed with the difference between the input and output magnitudes from the artificial stars tests (see Sect. \ref{sec:AST}).
 }
 \label{fig:CMD_NPL}
\end{figure}

The final CMD is shown in Fig.~\ref{fig:CMD_NPL}. Some distinctive features from foreground stars are present, like the disk Main Sequence (MS) stars at color $\mathrm{H-K} \sim 0.2$, and the disk RC stars. Most of the stars are heavily extincted. The RC stars are spread along a line from $\mathrm{H-K} \sim 1.0$ to $\sim 3.5$, which is often used to constrain the extinction (see section 3 for further discussion on this). Also, a prominent Asymptotic Giant Branch (AGB) bump is observed above the RC. The Red Giant Branch (RGB) and AGB sequences merge above the AGB bump, which is expected for these NIR filters. 

On the left part of the CMD we show typical errors for each bin of K magnitude. For the magnitude range where artificial stars were injected (see Sect.~\ref{sec:AST}), the black error bars represent the mean difference between injected and recovered magnitudes and colors within each K bin, as derived from the artificial stars tests (ATSs). For the remaining CMD region where no artificial stars were injected, we report the typical photometric uncertainty from \texttt{ALLFRAME}, shown in red. In both cases, the color uncertainty was computed by adding the H and K errors in quadrature. As expected, uncertainties increase towards fainter magnitudes, while the brightest bins show higher uncertainties due to their proximity to the saturation limit.

Sampling stars down to $\mathrm{K\sim20.5\pm0.3}$, the derived K, H-K CMD is one of the deepest obtained for the NB region from ground-based facilities. For comparison, the GNC reaches a magnitude limit of $\mathrm{K\sim18.5}$, which are discussed further in Sect.~\ref{sec:AST}. Other differences in the photometry of the GNC project are out of scope for this study.

\section{Kinematic determination of the extinction law} \label{sec:ext_law}

The extinction law towards the MW bulge has been derived by means of different techniques. \citet{nishiyama+2006} developed the "RC method", based on the fact that the intrinsic luminosity of RC stars changes little with metallicity and age \citep[e.g.,][for a comprehensive review]{girardi2016}; thus, they can be used as standard candles. They argue that if an RC is affected by different levels of extinction, it would change its position in the CMD along the reddening vector, with slope equal to the ratio of total to selective extinction $\mathrm{R_\lambda = A_\lambda / E_{\lambda'-\lambda}}$. Indeed, when we observe the CMD of the NBFs, which has severe differential extinction (see Fig.~\ref{fig:CMD_NPL}), the RC is spread along a line towards redder and fainter magnitudes. However, the underlying assumption is that the RC stars used for this derivation are all at the same distance, which is not necessarily true. While it is true that the NSD diameter ($\sim 400$~pc) is small compared to its mean distance, it is also true that foreground contamination from the disk and the bulge, at $\mathrm{b=0^\circ}$, is non-negligible and it might produce structures in the RC that modify its observed slope.

Focusing on the NB region, RC stars have also been used to study the extinction law. \citet{sanders+2022} combined data from VIRAC2 \citep{smith+2025}, GLIMPSE \citep{whitney+2011}, and the unWISE catalog \citep{schlafly+2019} to produce reddening maps for the inner regions of the Milky Way, by combining the RC method with the Rayleigh Jeans Color Excess method (RJCE). In order for their method to work, they assume that the distribution of RC stars peaks at the distance of the GC \citep{gravitycollab+2022}. Also, \citet{nogueras-lara+2020b} used the slope of the RC to measure the extinction law towards different regions in the GALACTICNUCLEUS survey, and later, \citet{nogueras-lara+2021c} used the derived values to create reddening maps using a two-layer approach.

To address the dust towards the NB, we developed a novel method to determine the extinction law based on the kinematics of the observed RC stars. We know that due to the rotation of the NB, stars in front of the NB must have different longitude proper motion ($\mathrm{\mu_l}$) from those behind the GC. Along with the assumption that the peak of RC density is located at the GC distance, we also assume that non-linear effects in the extinction are negligible. Thus, we measure where the position of the RC is when it has low but nonzero extinction, and compare it with the peak of the RC star distribution around the GC.

To implement this, we followed these steps: first, extract bona fide RC stars from the NBFs. Then, we must determine the position of the RC with negligible extinction (referred to as the "anchor RC"). Finally, we must find the RC stars near the GC using their kinematics (referred to as "kinematical center", KC0).

\subsection{RC Selection} \label{sec:RC_sel}

To select bona fide RC stars, we defined a parallelogram around this feature. To derive the slope of the upper and lower sides, we adopted the method called unsharp-masking technique \citep[see][]{De-Marchi+2016, Hosek+2018}. This method enhances the contrast in CMDs, making distinctive, high-density features such as the RC more prominent.

\begin{figure}
 \includegraphics[width=\linewidth]{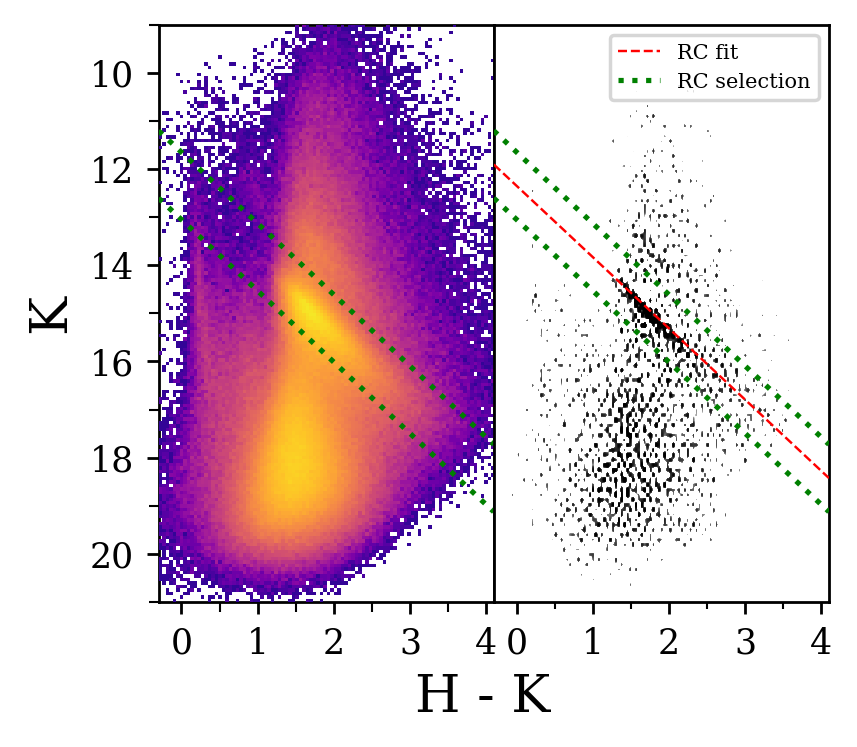}
 \caption{RC position using the unsharp masking technique. The green dotted line shows the upper and lower limits of RC selection. \textit{Left}: Hess diagram of our complete catalog. A logarithmic stretch was applied, in a range from 1 to 2\,000. Distinctive features are observed, such as the main sequence of foreground stars, the RC, and the AGB bump. \textit{Right}: Hess diagram after applying the unsharp masking technique. The RC is immediately recognizable. The thin dotted red line shows the best fit for the RC.}
 \label{fig:unsharp_masking}
\end{figure}

First, we created a Hess diagram binned by 0.1 in K and 0.06 mag in $\text{H - K}$, as seen in the left panel of Fig.~\ref{fig:unsharp_masking}. Then, we convolved the diagram with a 2D Gaussian profile of $\sigma=0.05$, to incorporate internal observational errors into the data. Afterward, we applied a second convolution with a wider Gaussian profile of $\sigma=0.2$ to create the "mask" that enhances high-frequency features. Finally, we subtracted the mask from the original Hess diagram, obtaining the one on the right of Fig.~\ref{fig:unsharp_masking}. As suggested in \citet{De-Marchi+2016}, we explored several values of $\sigma$ to create the mask, from 0.15 to 0.35, and we did not find any significant difference in the results.

Here, we applied a linear regression to compute the slope and the intercept that describe the peak RC stars. The resulting linear function is $K = 1.456(H-K) + 12.463$. In order to derive the extinction law, we define as RC sample the stars between $\pm$0.7 mag from this line.

\begin{figure*}[ht]
 \centering
 \includegraphics[width=0.95\textwidth]{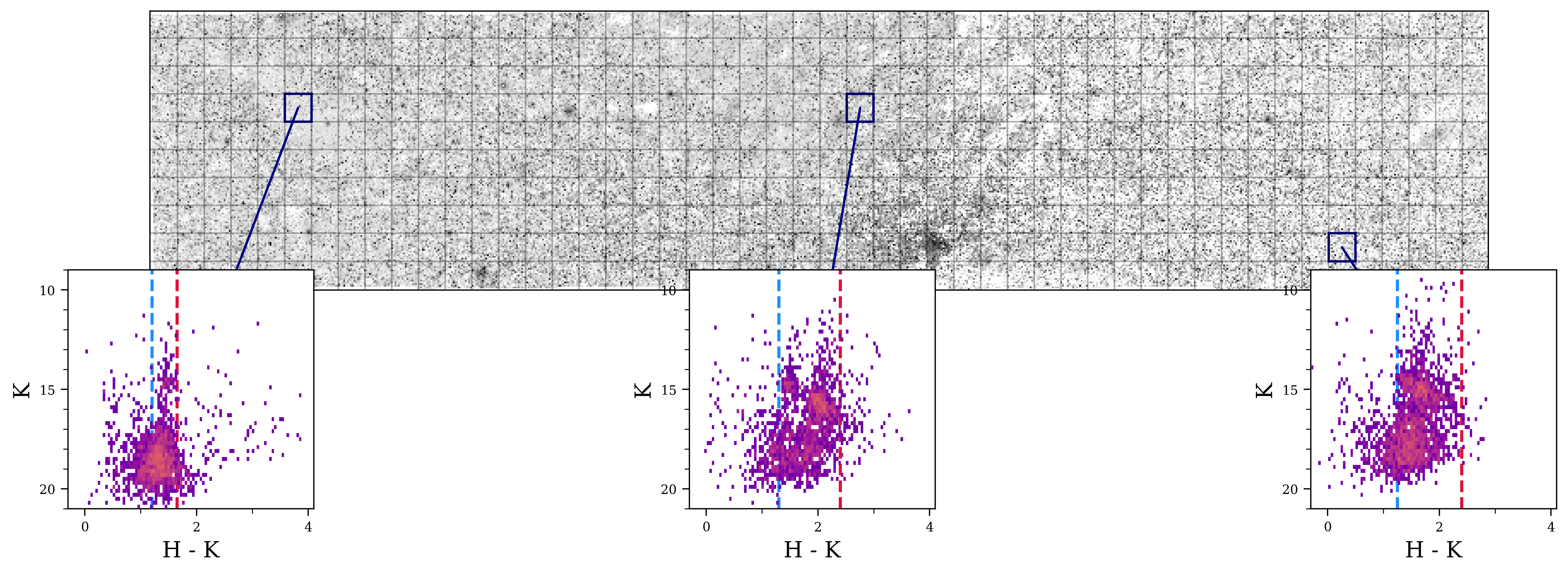}
 \caption{Same image as bottom panel of Fig.\ref{fig:nb_image}, showing examples of CMD in three different $50'' \times 50''$ regions. In all insets, the visually selected bluer and redder ends of each RC are shown in vertical blue and red dashed lines, respectively. The left inset shows a thin, well-defined RGB with a prominent RC in $\mathrm{H - K \sim 1}$. The middle inset is shows two RC features: one with low extinction at $\mathrm{H - K \sim 1}$, and another with high extinction centered at $\mathrm{H - K \sim 2}$. Between these two RCs, there is a gap without stars. The right inset shows a spread RC.}
 \label{fig:diff_ext}
\end{figure*}

The color limits of RC stars in the NB change dramatically, due to extinction, from one region to the next. This is illustrated in Fig.~\ref{fig:diff_ext} where we show the CMD of three regions $50'' \times 50''$ wide. For this reason, we selected, by visual inspection a blue and a red RC color cut in each spatial bin. The blue cut is usually evident, whereas the red one is more ambiguous, and therefore arbitrary, because the red side of the distribution falls off more smoothly. Nonetheless, the exact location of this cutoff does not affect the results presented here because we remove $\sim10$~\% of the stars, only in the red tail, not affecting the determination of the kinematic center (Sec.~\ref{sec:kinematical_center}) nor of the stellar density (Sec.~\ref{sec:stellar_mass}).

\subsection{Anchor RC position}\label{sec:anchor}

The next step is to determine what the observational position of the RC would be if the bulk of the stars were located at the GC distance and if the extinction were negligible. We decided to use the positions of three different low-extinction fields in the MW bulge to compute this point: Stanek ((l,b) = (0.25$^\circ$, -2.15$^\circ$)), SWEEPS ((l,b)~=~(1.25$^\circ$, -2.65$^\circ$)), and Baade's Window (BW, (l,b) = (1.04$^\circ$, -3.88$^\circ$)). To compute the position of the RC in these regions, we used VVV+VVVX PSF photometry from the catalog of Contreras Ramos (private communication) and selected a circular area with a radius of 0.5$^\circ$ around the central coordinates of each field. For consistency, we calibrated the photometry to 2MASS using the same procedure described in Sec. \ref{sec:data}. Then, we selected RC stars with $\mathrm{10 \le K \le 14}$. In this range of magnitude, we fitted the luminosity function (LF) with an exponential function and a Gaussian, using the least-squares minimization technique. We find the position of the RC as the mean of the Gaussian. Regarding the color in H$-$K, we employed a Gaussian mixture model to fit the data within the same magnitude range. We verified that the position of the RC in BW is consistent with that reported in \citet{zoccali+2021}, finding no significant offset in color nor magnitude.

We applied this method to the three fields. The positions of the RC are shown in the inset panel A of Fig.~\ref{fig:ExtLaw_AllFields}. The average is represented with a golden star, which is $\mathrm{(H - K, K) = (0.192, 13.135)}$. The difference in position of each RC along this window is small, less than 0.05 mag in color and magnitude.

\subsection{Kinematical center in the CMD}\label{sec:kinematical_center}

\begin{figure*}[ht!]
 \centering
 \includegraphics[width=0.80\textwidth]{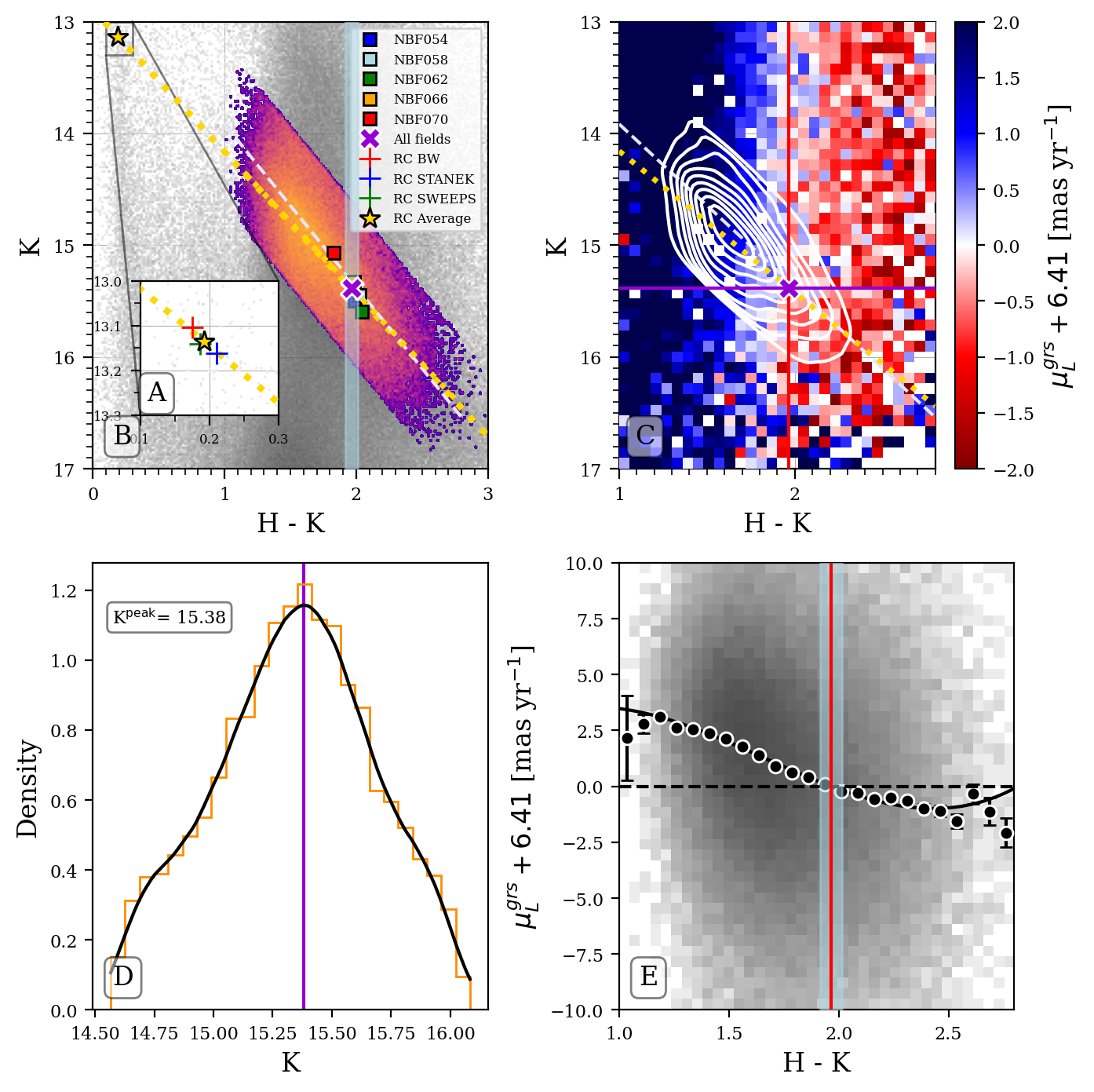}
 \caption{Determination of the extinction law. All the panels and inset share the same color coding and line style for the following: the best-fit for the RC (computed in Sec. \ref{sec:RC_sel}) is shown with a white dashed line; the derived extinction law is shown as a gold dotted line; the color where RC stars change the direction of rotation with respect to Sgr A* is represented with a red solid line and a shadowed gray area; and the K magnitude where the maximum of the RC is at the color of transition in kinematics is shown with a purple solid line. \textit{A:} Zoom in on the region near the RC anchor. Individual cross markers show the positions of the RC in low extinction windows, without uncertainties. The gold star with black edges represents the anchor RC position, which is the average of the RC in the windows. \textit{B:} CMD showing the RC selection colored by density. The gray dots represent all the stars not included in the RC selection. \textit{C:} CMD of NBF stars that cross-matched with VVV stars. Each bin is colored by the mean PM along the galactic latitude with respect to Sgr A*. The contours show the density of RC selection stars. \textit{D}: K LF within the 0.1 mag around $\mathrm{(H - K)^{KC0}}$. The orange histogram represents the LF, while the black solid line represents the KDE fit. \textit{E}: Longitudinal proper motions with respect to the proper motion of Sgr A* as a function of the observed color. The gray background represents the density of stars in this space. A logarithmic stretch was applied for better visualization. Black dots show the median of each 0.1 mag in color. The black solid line represents the best fit for the medians. The horizontal black dashed line shows where the proper motion with respect to the Sgr A* changes its rotation direction.}
 \label{fig:ExtLaw_AllFields}
\end{figure*}

To determine the KC0, we relied on the PM from the same PSF VVV+VVVX catalog. We did a cross-match with VVV and assigned those PM to all our NBF stars, which are calibrated to the Gaia Reference System ($\mathrm{\mu^{grs}}$). To find the kinematical center, we refer the PM with respect to Sgr A* $\mathrm{(\mu_l, \mu_b) = (-6.41, -0.219)~mas~yr^{-1}}$ \citep{reid&brunthaler2020}. 

Panel C of Fig.~\ref{fig:ExtLaw_AllFields} shows the CMD colored by the mean PM inside each bin. We detect a coherent change in the rotation of stars at color $\mathrm{H - K \sim 1.9}$. To find this color, we used panel E from Fig.~\ref{fig:ExtLaw_AllFields}, where we correlate the longitudinal PM with the color only for our RC stars that made a match with VVV. The gray squares in the background show the density of points. There is a clear trend with the color to change the direction of rotation. To model the rotation, we employed a running median algorithm with a bin size of 0.1 mag and fitted a third-degree polynomial to all the medians. The color where the rotation changes its direction is where the fit reaches zero, and that occurs at $\mathrm{(H - K)^{KC0} = 1.96}$.

We are left to find the magnitude of the peak RC stars around the kinematical center. To do so, we select all RC stars within 0.1 mag around $\mathrm{(H - K)^{KC0}}$ and find the maximum on their K LF. We used a linear kernel density estimator (KDE) from \texttt{sktlearn} to find the maximum in the LF and found a value of $\mathrm{K^{peak}} = 15.38$. This procedure is shown on panel D of Fig.~\ref{fig:ExtLaw_AllFields}.

Finally, we use the location of the KC0 in all fields (i.e. purple cross with white edges in Fig~\ref{fig:ExtLaw_AllFields}), with the position of the anchor RC (i.e., the golden star in Fig.~\ref{fig:ExtLaw_AllFields}, see Sec. \ref{sec:anchor}) to trace the extinction law. In panel A and B of Fig.~\ref{fig:ExtLaw_AllFields}, the derived extinction law is shown in a gold dashed line. The total-to-selective extinction ratio, and the H and K extinction ratios are:

\begin{equation}
    \qquad \mathrm{\frac{A_K}{E_{H-K}} = 1.259 \pm 0.074 \qquad \qquad \frac{A_H}{A_K} = 1.794 \pm 0.046}
\end{equation}

If our method is correct, then in each of the 5 NBFs we might measure a different KC0, because the GC might be seen behind different amounts of clouds, along different lines of sight, but all the KC0 should align, within the errors, along the same extinction vector derived above. Indeed, we verified our results by measuring again the total-to-selective extinction ratio, this time performing a least-squares fit using all NBFs points (i.e., squares in Fig~\ref{fig:ExtLaw_AllFields}, panel B), and forcing the fit to pass through the RC anchor position (i.e., the golden star in Fig.~\ref{fig:ExtLaw_AllFields}, panels A and B). Since the RC locus is expected to move along the extinction vector as the extinction varies \citep{nishiyama+2006, nishiyama+2009}, it is normal to have some KC0 in some fields less extincted than others. In fact, the less extincted field is the NBF070, towards the LOS with less dust in the CMZ. Using this method, we obtained a total-to-selective extinction ratio of $\mathrm{A_K/{E_{H-K}} = 1.265}$, and an extinction ratio of $\mathrm{A_H/A_K = 1.791}$, which are consistent with the values computed using all the fields. 

We estimated the uncertainty in the total-to-selective extinction ratio by resampling the KC0 positions of each individual NBF field from normal distributions, and repeating the linear fit 20~000 times, always forcing it to go through the RC anchor position. We computed the uncertainty as the standard deviation of the distribution of total-to-selective ratios. We considered homoscedastic errors of 0.1 mag in color, corresponding to the bin width used to find the kinematic center for each field. Whereas for the K magnitude, we used 0.25~mag, which is the approximate width of the K LF on each field. To estimate the uncertainty in the absolute extinction ratios, we analytically propagated the errors.

All the individual KC0 and the one using all fields are shown in panel B of Fig.~\ref{fig:ExtLaw_AllFields}. As observed, each individual field is scattered around the extinction law derived using all fields, which could suggest a dependency of the extinction law variation with the LOS. These spatial variations have been observed in the NIR regime in previous works, but on larger scales \citep[e.g.][]{nishiyama+2006, alonso-garcia+2017}. Overall, our derived extinction ratios are consistent with previous measurements in the literature towards this region of the MW, as discussed in Sect.~\ref{sec:ext_comparison}.

\subsection{High-resolution reddening map}

\begin{figure*}[ht]
 \centering
 \includegraphics[width=0.8\textwidth]{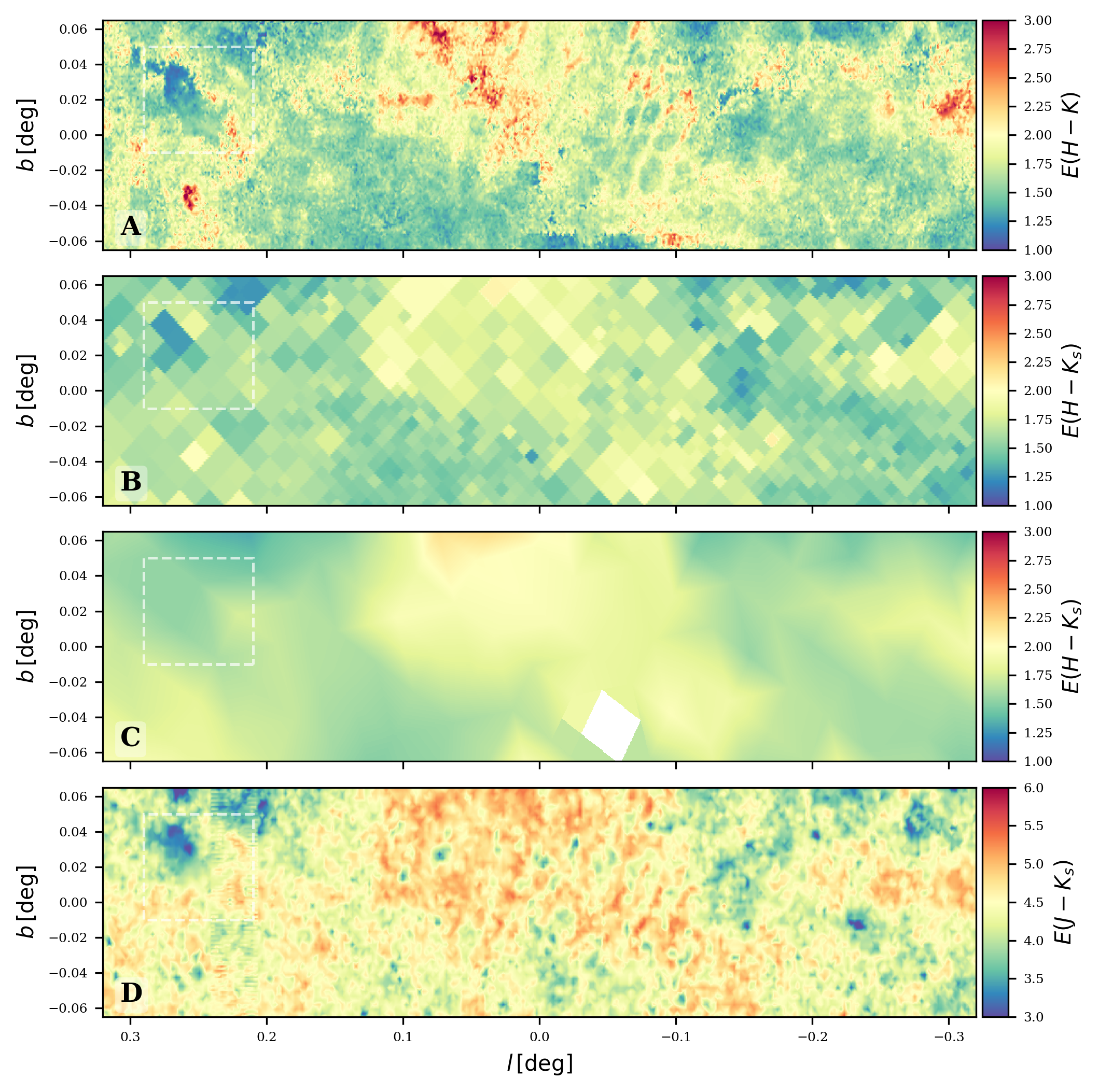}
 \caption{Reddening maps available for the NB region from different authors. \textit{A}: This work. \textit{B:} \citet{sanders+2022}. \textit{C}: \citet{zelakiewicz+2025}. \textit{D:} \citet{surot+2020}. Panel D map uses J instead of H. In all panels, a dimmed white-dashed rectangle shows the location of the Brick, we show later in Fig.~\ref{fig:brick+gas}.}
 \label{fig:ExtinctionMaps}
\end{figure*}

With this total-to-selective extinction ratio and our selection of RC stars, we constructed a reddening map for the region. First, we binned our region with a bin size of $4.5'' \times 4.5''$. With these bins, we have a mean of five stars RC stars per bin. For each RC star, we compute their reddening assuming a reference color of $\mathrm{(H-K)_0 = 0.1}$, which is the observed intrinsic color in the same region \citep{nogueras-lara+2021c}. We average the extinction of the RC stars within each spatial bin.

This initial map has some bins with no stars. To fill them, we reconstructed the map with the \texttt{astrofix} package \citep{zhang&brandt2021}. In short, this package interpolates images using a Gaussian Process Regression to fix bad pixels. When applied to our reddening map, it predicts the extinction value for each bin\footnote{\texttt{astrofix} is intended to fix astronomical images using this exact same algorithm, but it is also useful for this science case.}. We are confident in applying this method to our map because only $\sim5\%$ was fixed using this method. It is worth mentioning that in bins with no RC stars could be due to strong interstellar extinction. If that is the case, the amount of extinction in the interpolated bins must be higher than in their surroundings. Therefore, these bins should be considered as a lower limit for the extinction.

Finally, we assigned an extinction value to each star in the NBFs interpolating the binned reddening map using a Radial Basis Function (RBF) interpolator implemented in \texttt{scipy} \citep{SciPy-NMeth2020}. We used a thin-plate spline kernel, with a first-degree polynomial added to reproduce the linear trends across the field. We interpolated using the 10 nearest bin centers to each query point, ensuring that each star's assigned extinction value reflects the local structure of the map rather than a global fit. The result of this map can be observed on panel A of Fig.~\ref{fig:ExtinctionMaps}. In the same figure, we show the reddening maps from \citet{sanders+2022}, \citet{zelakiewicz+2025}, and \citet{surot+2020} on panels B, C, and D, respectively.

\begin{figure}[ht]
 \includegraphics[width=\linewidth]{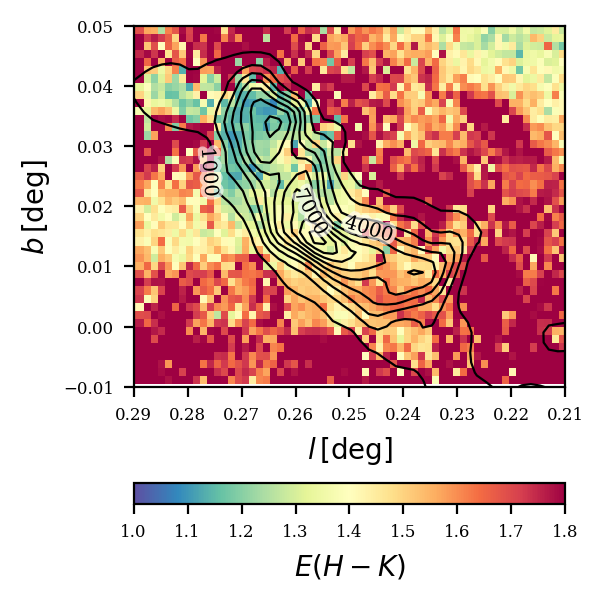}
 \caption{Zoom in of Fig~\ref{fig:ExtinctionMaps} of the Brick region. The colorbar is optimized to highlight the gradient in $\mathrm{E(H-K)}$ over the Brick. The northern part of the brick is less extincted than the southern one. The contours show the column density maps from \citet{marsh+2017}, in units of $10^{20}~\mathrm{cm^{-2}}$.}
 \label{fig:brick+gas}
\end{figure}

Some distinctive features are evident when observing our map. First, there is a clear gradient in the reddening on top of the Brick ($\mathrm{(l,b)\sim(0.25^\circ,0.02^\circ)}$). We can observe the location of the Brick in Fig~\ref{fig:ExtinctionMaps} displayed by white-dashed lines, and the zoom in of our reddening map on top of the Brick in Fig~\ref{fig:brick+gas}. The northern region of the Brick is less extinct than the southern part. This is not an artifact: the gradient is also evident in the map of \citet{sanders+2022}, where the less reddened region is visible, although the spatial resolution does not allow a clear distinction of the gradient. On the other hand, the map of \citet{surot+2020} clearly shows the less-extinct region, but the higher-extinction region merges with the rest of the map. We interpret this as they used J instead of H; therefore, they are limited by this bluer band, for which extinction is $\sim$2 times higher than for H, and $\sim$3 times higher than for K. The low extinction on top of the Brick is less clear in the map by \citet{zelakiewicz+2025} than in the other maps. Additionally, all the features on their map are less defined than those on ours. This is mainly due to the spatial resolution of HAWK-I compared to the instruments used to create the other maps. Moreover, the shape of the Brick we observe in IR extinction matches the overall emission of some dense gas tracers \citep[][their Fig. 1]{rathborne+2014} and continuum emission \citep[contours in Fig.~\ref{fig:brick+gas} from][]{marsh+2017}.

\section{Stellar mass estimation through RC star counts}\label{sec:stellar_mass}

One way to estimate the stellar mass is through RC star counts, as described in \citet{valenti+2016}. Computing the mass on each NBF leads us to improve the mass profile shown in their Fig. 4, by adding five points instead of one in their most central bin, where all NBFs are located (see inset of Fig.~\ref{fig:RC_density_profile}). To achieve this goal, we performed the following steps: first, we corrected the observed star counts for completeness. Then, we deredden the K LF and subtracted the RGB stars' contribution. Finally, converted the number of RC stars to mass using a fully empirical approach.

\subsection{Artificial Star Tests} \label{sec:AST}

\begin{figure}[ht!]
 \includegraphics[width=\linewidth]{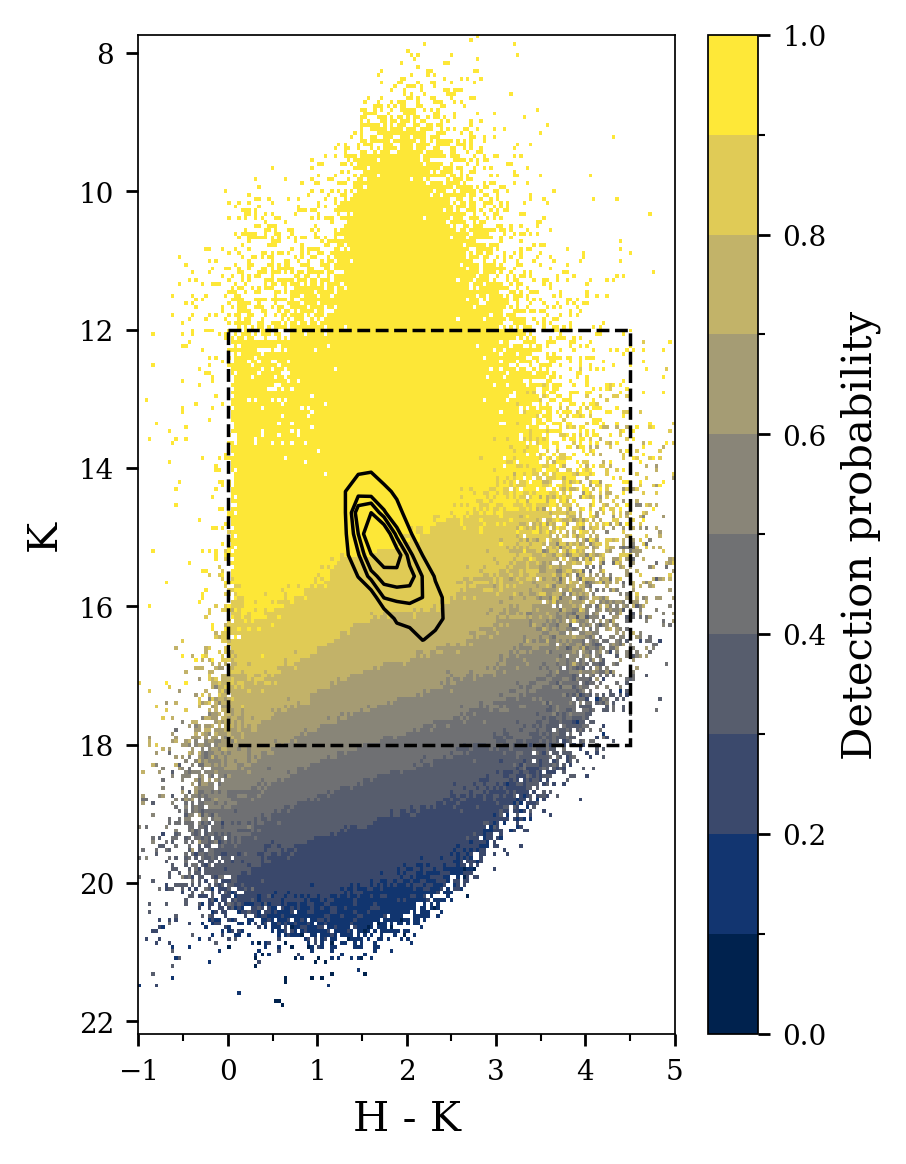}
 \caption{Same Hess diagram as Fig.~\ref{fig:unsharp_masking}, but colored by detection probability. The injection region is shown in the black dashed line rectangle, between $12~\le~\text{K}\le~18$ and $0~\le~\text{H-K}~\le~4.5$. The black contours show the quintile levels of the RC density.}
 \label{fig:CMD_prob-dect}
\end{figure}

We estimated the probability of each star being detected given its position in the sky and locus in the CMD by performing artificial star tests (ASTs). This is a classical technique widely used in the literature \citep[see e.g.,][Sec. 2 for a review]{aparicio&gallart1995, gallart+1996, harris&speagle2024}. Because the RC spans a significant range both in magnitude and in color due to extinction, stars were injected within a CMD box encompassing the whole RC region (Fig.~\ref{fig:CMD_prob-dect}) very generously. A flat distribution was used for K magnitude and H-K color.

The stars were injected in a hexagonal grid designed to avoid artificial crowding \citep[see][]{piotto+99, zoccali+1999, zoccali+2003}. We used 40 pixels of separation between artificial stars ($\sim$ 2 PSF fitting radius), which allowed us to inject $\sim 3\,000$ stars simultaneously per chip per iteration. We repeated the process 31 times to get $\sim 2\,000\,000$ artificial stars in all the NBFs.

To estimate the completeness, we use the recovery fraction, which is the ratio between recovered and injected stars at a given locus in the CMD \citep{aparicio&gallart1995}. To assess the different crowding levels across the images, we created Voronoi bins where the crowding is approximately the same. Within each bin, we modeled the completeness as a logistic function and fitted it to the observed recovery fraction in color and magnitude at the same time. With this model, we assigned detection probabilities to each real star. Stars outside the box were extrapolated from the model. 

In Fig.~\ref{fig:CMD_prob-dect}, we show the CMD of all NBF fields colored by detection probability. As expected, the brighter, less extincted stars have higher detection probabilities than the fainter, more extincted ones. Also, we verify that RC stars are in a high-completeness region. The bulk of RC stars are located in the 90\% completeness region, whereas the most extincted RC stars have between 70\% and 80\% completeness. Also, we check that the 50\% probability detection level is at $\mathrm{K = 18.0 \pm 0.3}$, $\mathrm{H-K = 2.0 \pm 0.3}$, which is 3 magnitudes fainter than the peak of the observed RC. This is an improvement over previous completeness estimates in this region. \citet{valenti+2016} in their Fig. 1 show a completeness below 50\% for this region at the dereddened magnitude of the RC, obtained with AST as well. The 50~\% completeness threshold is important because one way to define the limiting magnitude \citep[e.g., ][]{harris1990}. Also, \citet{nogueras-lara+2020b} estimated a completeness of 80\% at magnitude \Ks\ $\sim 16$, which is somewhat smaller than our estimations of completeness at the same magnitude. However, they did not use AST to estimate completeness, which, along with GNC's different observation strategy, could explain this difference.

\subsection{Stellar density profile}

In this Section, we derived the stellar density profile based on the dereddened LF of RC stars. We adopted an intrinsic color for the RC of $\mathrm{(H-K)_0 = 0.114}$. This is the color of the chosen anchor point (see Sect. \ref{sec:anchor}), hence it is not exactly the RC intrinsic color, as our reference fields do have some reddening. Nonetheless, this value is only slightly larger than the intrinsic color adopted by \citet[e.g.][]{plevne+2020, nogueras-lara+2020b, nogueras-lara+2021}, as expected, given that the chosen reference fields are all low reddening windows.

\begin{figure*}[ht!]
\centering
 \includegraphics[width=0.85\textwidth]{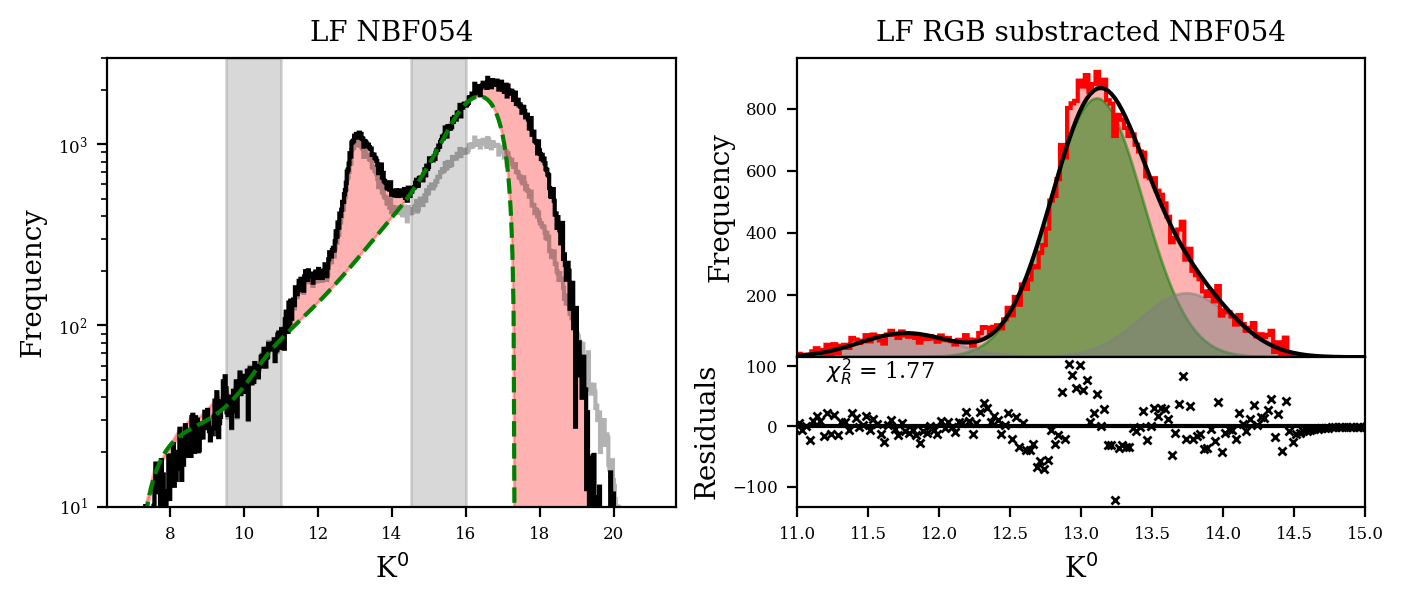}
 \caption{Star counts of field NBF054. \textit{Left}: \Ko LF of field NBF054. The black line shows the star counts corrected by completeness. For comparison, the gray line represents the uncorrected LF. The spline modeling the RGB is shown in green dashed lines. The fitting regions for the spline are shown as a vertical gray area. The LF with subtracted RGB LF is represented with a red area. \textit{Right}: The \Ko\ LF with RGB subtracted is represented by the solid red line histogram. There are three Gaussians fitted to the data: in gray, the left and right ones represent the early-AGB and RGB-bump, respectively; whereas the green central one represents the RC star counts. The black solid line represents the sum of the three Gaussians. Residuals to the fit are shown in the bottom part of the panel.}
 \label{fig:LF_fit_npl054}
\end{figure*}

On each NBF field, we corrected the observed star counts by weighing the dereddened LF by the inverse of the detection probability, obtained from the completeness analysis. To get the number of RC stars, we need to discount the contribution of RGB stars in the LF, as can be seen in the left panel of Fig.~\ref{fig:LF_fit_npl054}. We fitted the RGB with a cubic spline in the areas next to the RC in the LF, depicted as the gray areas if Fig \ref{fig:LF_fit_npl054}. Next, we substracted the spline to the LF and obtained the histogram shown in the right panel of Fig.~\ref{fig:LF_fit_npl054}. Then, we used the \texttt{BayesianGaussianMixture} algorithm from \texttt{scikit-learn} to find the best composition of Gaussian functions that describe the LF with the RGB substracted. In all the NBFs, we find that the data is well described with three distributions, which we interpret as the early-AGB phase, the RC stars, and the RGB-bump. Also, we note the fit shows a peak in the residuals at \Ko~$\sim 13$. This peak could be due to the shape of the RC in the LF not being perfectly described by a Gaussian. Nevertheless, the residuals are symmetric around zero, with no systematic trend, which indicates the fit is unbiased.

\begin{figure}[ht!]
 \includegraphics[width=\linewidth]{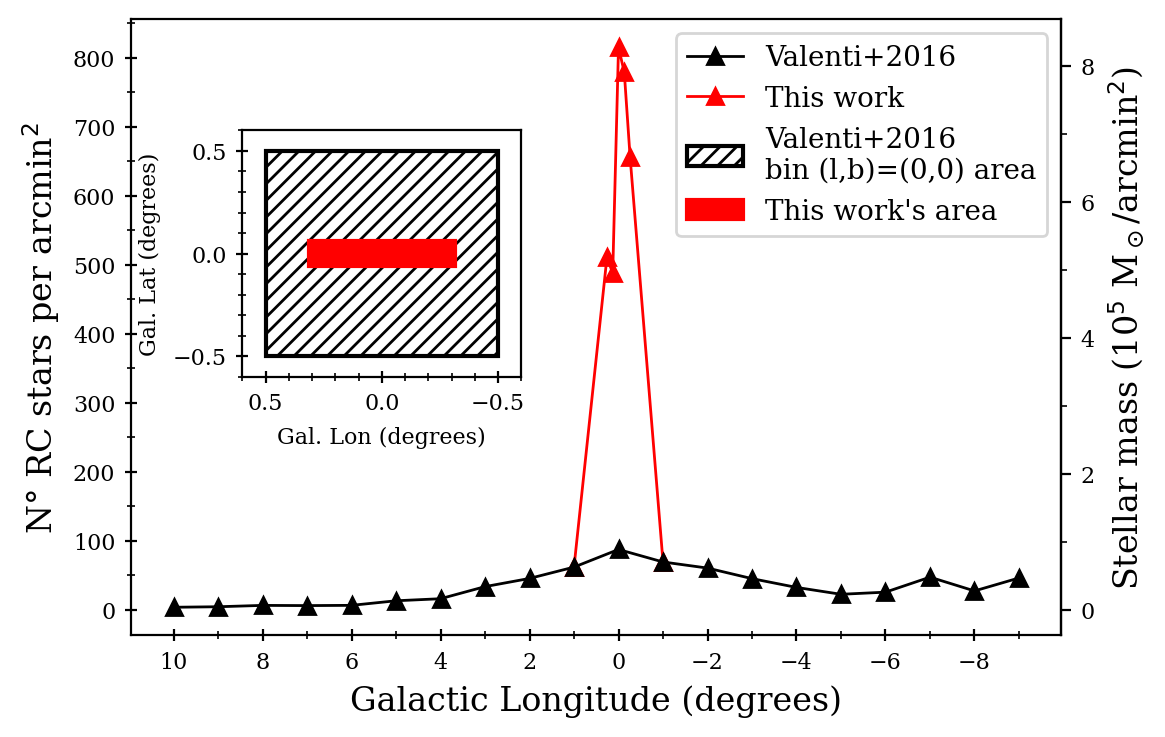}
 \caption{Stellar density profile, in number of stars per $\text{arcmin}^2$ as function of the galactic longitude. The black triangles show the density profile computed by \citet{valenti+2016} in their Fig. 4 at $b = 0^\circ$. The red triangles represent the 5 fields studied in this work. The inset shows a comparison between the spatial extent of both works. The hatched-black square represents the innermost bin of \citet{valenti+2016}, whereas the filled red rectangle is the area covered in this study. The right axis shows the star counts converted to stellar mass per arcmin$^2$. The difference in the spatial coverage explains the marked difference in RC star density.}
 \label{fig:RC_density_profile}
\end{figure}

Fig.~\ref{fig:RC_density_profile} compares star count results with the ones of \citet[][their Fig. 4]{valenti+2016}. To do so, we normalize the counts to the same unit of area. We notice that our area is much smaller than the VVV area: in fact, all of our fields lie within their innermost field, as shown by the inset of Fig.~\ref{fig:RC_density_profile}. In fact, the purpose of this Section is to provide a density and a mass profile with higher spatial resolution in the NB of the MW, where the spatial gradient is very large, and an average of $1^\circ\ \times 1^\circ$ field cannot appropriately describe the real profile. The innermost point of the profile of \citet{valenti+2016} is the mean of the high density of the NB with much lower density of a larger area surrounding it. Obviously, it is a factor of 10 lower than the real density of the Nuclear region, which we provide with our finer spatial sampling in the inner region.

\subsection{Estimation of the stellar mass}

To convert RC star counts into stellar mass, we rely on the measurements by \citet{valenti+2016} of the mass of the MW Bulge. Their method employs a three-step scaling approach that begins with the empirical IMF derived from HST/NICMOS observations in a small bulge field \citep{zoccali+2000, calamida+2015}. Then, it is scaled up to a medium-sized NTT/SOFI, as discussed in \citet[][their Sec. 5]{zoccali+2003} using the area ratio between the two observations, and finally extrapolated to the entire VVV survey area by comparing RGB and RC star counts between the SOFI field and the full VVV footprint \citep{valenti+2016}. 

\citet{valenti+2016} found a total mass of the MW bulge of $\mathrm{2.0\times10^{10}~M_{\odot}}$. Their bin centered in $\mathrm{(l,b) = (0^\circ, 0^\circ)}$ has a mass of $\mathrm{3.2\times10^{8}~M_{\odot}}$ (their Fig. 5) with 315\,274 RC stars (private communication and their Fig. 4), meaning that each RC star traces $1\,015~\mathrm{M}_\odot$ of stellar mass. On the other hand, in all the NBFs, we find a total of 182\,729 RC stars, thus tracing a total stellar mass of $\mathrm{1.85 \pm 0.3 \times 10^{8}~M_{\odot}}$ within the region studied here. Since our data only covers the innermost part of the NB, we scaled up the observed mass by considering the ratio between the observed area of NBFs and the NSD extent as reported by \citet{sormani+2022}. We considered a box defined by their radial scale length and scale height as the extent of the NSD model, with values of 88.6~pc and 28.4~pc, respectively. This gives an area ratio of $\sim6.6$ between the full NSD model and the NBF footprint. We assumed a uniform distribution of stellar mass across these areas, which is a simplification of the density gradient of this structure. With this scale factor, our scaled mass of the NB is $\mathrm{12.2 \pm 2.6\times 10^{8}~M_{\odot}}$. The uncertainties of the stellar mass, and a comparison of our measurement with model predictions and literature estimates are presented in Sect.~\ref{sec:mass_discussion}.

\section{Discussion}\label{sec:discussion}

\subsection{The distance-extinction degeneracy in the NBFs}\label{sec:degeneracy}

There is a difference between the extinction ratios computed with our kinematic method, and those derived from a direct linear fit to the RC in the NBFs. The linear fit produces a slope of 1.456, which differs from our total-to-selective extinction ratio of $1.256\pm0.076$ by more than $2\sigma$. This discrepancy comes from a distance-extinction degeneracy that affects the observed RC slope in the NBFs. As shown in Fig~\ref{fig:ExtLaw_AllFields} panels C and E, less-extincted stars have systematically different proper motions than more-extincted stars, indicating that they are not at the same distance. Stars with lower extinction are located preferentially in the foreground, whereas stars with high extinction are located preferentially in the background. Hence, the observed spread in K magnitude along the observed RC sequence arises from a superposition of increasing extinction and distance modulus, leading the RC slope to overestimate the total-to-selective extinction ratio.

We verify this quantitatively. By adopting $\mathrm{A_K/E(H-K)~=~1.456}$ and an intrinsic color of $\mathrm{(H-K)_0 = 0.1}$, the dereddened RC locus results at $\mathrm{(H-K,K)~=~(0.1,12.61)}$. Then, assuming an absolute magnitude of $\mathrm{M_K~=-~1.60}$, we get a distance of $\sim6.9~\mathrm{kpc}$ to the bulk of the RC stars, which is significantly less than the well-established distance to the GC. By contrast, forcing the fit to pass through the anchor RC position breaks this degeneracy and produces a physically consistent slope of $\mathrm{A_K/E(H-K)~=~1.265}$. Since the distance to the GC is well established \citep[e.g.,][]{gravitycollab+2022}, we rely on the anchor RC as a robust constraint on the fit.

One could alternatively interpret these results as the bulk of RC stars being located in front of the GC, which lacks a physical justification. There is no reason to expect a concentration of RC stars at $\sim6.9~\mathrm{kpc}$ along this LOS. We therefore conclude that our method breaks the distance-extinction degeneracy, and it recovers the intrinsic extinction law.

\subsection{Comparison with previous extinction ratios}\label{sec:ext_comparison}

\begin{table}[]
    \caption{Absolute and total-to-selective extinction ratios from previous works.}
    \centering
    \begin{tabular}{c c c}
        \hline\hline
        $\mathrm{A_H / A_{K}}$ & $\mathrm{A_{K} / E_{H-K}}$ & References \\
        \hline
        1.79 & 1.259 & This work \\
        --   & 1.87  & 1 \\
        1.55 & --    & 2 \\
        1.73 & 1.44  & 3 \\
        1.73 & 1.61  & 4 \\
        1.76 & --    & 5 \\
        --   & 1.28  & 6 \\
        1.88 & 1.104 & 7 \\
        1.99 & --    & 8 \\
        1.84 & 1.22  & 9 \\
        1.76 & 1.30  & 10 \\
        1.77 & 1.293 & 11 \\
        --   & 1.320 & 12 \\
        \hline\hline
    \end{tabular}
    \tablebib{(1)~\citet{cardelli+1989}; (2)~\citet{indebetouw+2005}; (3)~\citet{nishiyama+2006}; (4)~\citet{nishiyama+2009}; (5)~\citet{schoedel+2010}; (6)~\citet{alonso-garcia+2015}; (7)~\citet{alonso-garcia+2017}; (8)~\citet{Hosek+2018}; (9)~\citet{nogueras-lara+2020b}; (10)~\citet{minniti+2020}; (11)~\citet{sanders+2022}; (12)~\citet{albarracin+2025}
    }
    \label{tab:ext_laws}
\end{table}

Our ratios are in good agreement with works that used RC stars to trace the extinction law in the GC region. \citet{schoedel+2010} measured the total extinction in RC stars by adopting an absolute magnitude of the RC stars $\mathrm{M_K = -1.54}$ \citep{groenewegen2008}, a mean distance to the GC of $\mathrm{R_0 = 8.03~kpc}$, and an intrinsic color of $\mathrm{H-K} = 0.07$. With this, they estimate an absolute extinction ratio of $\mathrm{A_H/A_K = 1.76}$. One key point is that \citet{schoedel+2010} used only stars within a FoV of $40'' \times 40''$ around Sgr A*, whereas our spatial coverage is $\sim630$ times larger than theirs. Despite the difference in spatial coverage, the results are consistent, meaning that the extinction properties near the GC are compatible with the extended ones of the NB.

With a wider spatial coverage, \citet{nogueras-lara+2020b} computed extinction ratios using the RC method in the CMD of different regions of the GNC project, in J, H, and Ks filters. For their central region, which is the one with the most overlap with our NBFs fields, they find an extinction index of $\mathrm{\alpha_{HK} = 2.20}$, which we converted to the total-to-selective extinction ratio using their Equation 1, obtaining $\mathrm{A_K/{E_{H-K}} = 1.22}$. Also, they report the absolute extinction ratio of $\mathrm{A_H/A_K = 1.84}$. It is noticeable that their RC slope differs from the slope we measured in Sec. \ref{sec:RC_sel}. To double-check our results, we measured the slope using the same method as \citet{nogueras-lara+2020b}, by binning by color and applying a Gaussian Mixture Model on each bin. We find a RC slope of 1.44, consistent with the measurement of the unsharp-masking method. Although the methods used are very different, the extinction ratios are similar.

Interestingly, our kinematic derivation of the extinction ratios is consistent with other methods and tracers. \citet{sanders+2022} used a multi-wavelength approach using the RC method plus the Rayleigh-Jeans Color Excess method in the inner $3^\circ \times 3^\circ$ of the Galaxy. Their measurements of the total-to-selective extinction ratio and absolute extinction ratio are very similar to our independent measurement, with values of $\mathrm{A_K/{E_{H-K}} = 1.293}$ and $\mathrm{A_H/A_K = 1.77}$, respectively. Also, our results are in agreement with methods that used variable stars as tracers. \citet{minniti+2020} used Classical Cepheids to trace the disk structure on the far side of the MW bulge, and they also derived extinction ratios by measuring the color excess ratios and knowing the intrinsic colors of these variable stars. They obtained $\mathrm{A_K/{E_{H-K}} = 1.30}$ and $\mathrm{A_H/A_K = 1.76}$. Finally, \citet{albarracin+2025} did a similar work with Mira variables, and found $\mathrm{A_K/{E_{H-K}} = 1.32}$. All of these results point towards a convergence on the values of extinction ratios towards the NB of the MW.

Our results differ from the classical extinction ratios assumed for the rest of the Galaxy. The total-to-selective extinction ratio from \citet{nishiyama+2006, nishiyama+2009} are $\mathrm{A_K/{E_{H-K}} = 1.44}$ and $\mathrm{A_K/{E_{H-K}} = 1.61}$ respectively. They were computed using the RC method, but in the outer parts of the MW bulge. Also, \citet{alonso-garcia+2017} used the same method in the VVV survey filters, and found $\mathrm{A_K/{E_{H-K}} = 1.104}$ for the bulge. As they state in their work, this value varies depending on the selected LOS. 

We present a summary of the works we compared in Table \ref{tab:ext_laws}. The extinction ratios are stable across different methods and tracers, suggesting they are not strongly sensitive to the specific technique employed. Therefore, we consider our derived extinction ratios robust for this region of the MW.

\subsection{Assumptions of the RC method}\label{sec:assumptions}

The RC method has some underlying assumptions that we need to address for the NB. First, it assumes that all stars are at the same distance, or at least the spread along the LOS is negligible compared with the distance to the object of study. This assumption is true for studies of satellite galaxies (i.e., the LMC and SMC), but not for the NB. Here, the expected LOS spread is about 400~pc, which represents a variation of $\sim 0.1$ magnitudes at 8.2~kpc. If the spread along the LOS is not carefully addressed, it could be misinterpreted as population effects.

This leads us to a second assumption: RC stars' intrinsic brightness varies little with age and metallicity distribution. This is true for stars older than 1 Gyr \citep[see][]{girardi2016}, and underpins both the RC method and our kinematic selection of RC stars to measure the extinction law. In the NB, however, we expect a spread in both metallicity \citep[][, their Fig. 10]{fritz+2021} and age. The latter is expected due to the inside-out formation observed in extragalactic nuclear regions \citep{Bittner+2020}, and is detected in the MW NSD by \citet{nogueras-lara+2023}. On this matter, \citet{nogueras-lara+2020a} concluded that about 90\% of the stellar mass of the NSD was formed over 8 Gyr ago; 5\% was formed about 1 Gyr ago, and the remaining 5\% is less than 0.1 Gyr old. To reach this conclusion, they decomposed the GNC de-reddened, completeness-corrected \Ks\ LF as a linear combination of theoretical LFs for different stellar populations. Although simple, this method can be subject to up to 20\% error if systematic effects are not considered \citep{girardi2016}.

While the RC method is robust to moderate variations in these parameters, we must caution that the distribution of star ages in the NB remains uncertain. This could introduce errors in our measurements, particularly in the interpretation that the peak of the LF around $\mathrm{(H-K)^{KC0}}$ corresponds to stars at the distance of the GC (see Sect. \ref{sec:kinematical_center}).

\subsection{Interpretation of RC distributions in the NB}\label{sec:brick}

The RC distribution in the NB shows structures that are not completely understood, and their interpretation demands caution.

For example, \citet{zoccali+2021} questioned the distance used for the Brick. That study found the RC on top of the Brick is significantly brighter than an RC in a control field. This vertical offset was interpreted as a distance difference, placing the Brick at 7.2 kpc from the Sun. Later, \citet{nogueras-lara+2021} computed the distance using two methods: 1) a linear fit to the observed RC to recover the mean reddening-free \Ks\ magnitude, and 2) by fitting the de-reddened RC LF to obtain the mean distance to the Brick. With the former method, they obtained a distance of 7.6 kpc, whereas using the latter, they got 8.4 kpc. Additionally, \citet{lipman+2025} computed a probability of a group of dense clouds located in the direction of the NB to be on the near side or on the far side of the NB. They did it by using the 8$\mu m$ and 70 $\mu m$ dust extinction method. To do so, they used GLIMPSE residual map from Spitzer telescope \citep{Benjamin+2003, churchwell+2009}, and the Hi-GAL \citep{molinari+2010, molinari+2016} 70 $\mu m$ emission and the computed column densities. They obtained that the Brick is most likely lying on the near side of the NB.

Our map shows that the mean reddening above the Brick is lower than that of its surroundings, which supports the hypothesis that the Brick lies in front of the NB. However, there is no gradient in gas density or continuum emission along the major axis of the Brick, which cannot explain the gradient observed in IR extinction, as observed in Fig.~\ref{fig:brick+gas}. Although the Brick being in the NB is a reasonable assumption, the differences in results across methods point to an underlying problem in interpreting the RC distribution and a complex 3D structure of the Brick \citep{henshaw+2019}.

Furthermore, the LF distribution in Fig.~\ref{fig:LF_fit_npl054} shows three components, 2 of which are clearly the AGB-bump and RC. However, the interpretation of the third, fainter component is not clear. In this work, we interpret this faint component as a metal-rich RGB-bump, since for the observed metallicities in the NB it should fall below the RC \citep[e.g.,][]{valenti+2004b}. However, \citet{nogueras-lara+2020a} interpret the same feature in their data as a secondary RC, which they associate with a younger population in the NB. The ambiguity between these two interpretations highlights the complexity of the stellar populations in the NB, and underscores the need for a broader parameter space to disentangle its different components.

\subsection{Stellar mass comparison with models and literature}\label{sec:mass_discussion}

To compare with model predictions over the same spatial footprint, we used \texttt{AGAMA} software \citep{vasiliev2019} to evaluate the stellar mass density of \citet{sormani+2022} and \citet{hunter+2024} models. We drew $10^6$ random points uniformly distributed within a volume defined by the NBF longitude and latitude, and a distance range of 7--9~kpc. The distance range was chosen to enclose the NSD, which dominates the stellar mass along this LOS, and to include any foreground and background contamination. We computed the enclosed mass as the product of the mean model density and the sampled volume. For \citet{sormani+2022} we find $\mathrm{0.73 \times 10^{8}~M_{\odot}}$, whereas for \citet{hunter+2024} we find $\mathrm{1.63 \times 10^{8}~M_{\odot}}$. Our observed mass of $\mathrm{1.85 \times 10^{8}~M_{\odot}}$ is consistent with predictions from dynamical models for this region of the MW.

The uncertainty of the stellar mass estimate has two main contributions. The first comes from the mass traced by each RC star, for which we adopt the 15\% conservative uncertainty reported by \citet{valenti+2016}, which accounts for the IMF slope, disc contamination and stellar population gradients. This yields an observed stellar mass of $\mathrm{1.85 \pm 0.30 \times 10^{8}~M_{\odot}}$. This error dominates over other sources of uncertainty, such as the Poisson error when counting RC stars, or the uncertainty in the RGB subtraction from the \Ko~LF.

The second contribution of uncertainty comes from the area scaling, which assumes a uniform surface density across the NSD model. Since the NBF samples the densest part of the NSD model, this assumption likely overestimates the total mass, and introduce an additional systematic uncertainty, that we conservatively estimate at the same level of the first: 15\%. Applying both contributions to the scaled mass, and combining them in quadrature, we obtain a total uncertainty of $\mathrm{\Delta M = 2.6 \times 10^{8}~M_{\odot}}$, and a final scaled mass of $\mathrm{12.2 \pm 2.6 \times 10^{8}~M_{\odot}}$ for the NB region.

This value is in agreement with the reported mass estimates in the NB, as shown in Table~\ref{tab:nb_mass_comp}. \citet{sormani+2020b} found a dynamical mass of $\mathrm{6.9 \pm 2 \times 10^{8}~M_{\odot}}$ applying a Jeans modelling to the NSD region. \citet{li+2022} used this value in hydrodynamical simulations with updated potentials for the region (see their Table 1), and verified that an NSD with this stellar mass can reproduce a gas disk similar to the observed CMZ. Later, \citet{sormani+2022} updated their result to $\mathrm{10.5 \pm 1 \times 10^{8}~M_{\odot}}$ with a self-consistent dynamical model. Even earlier than these measurements, \citet{launhardt+2002} estimated a mass of $\mathrm{14.2 \pm 6 \times 10^{8}~M_{\odot}}$ in the NB region, by assuming a fixed mass-to-light ratio. It is worth noting that this former estimate and ours do not rely on dynamical models for the mass estimation in the NB. Our empirical measurement is consistent with these results within the reported errors.

\begin{table*}[t]
    \caption{Mass estimates in the Nuclear Bulge region compared with literature.}
    \centering
    \setlength{\tabcolsep}{5pt}
    \renewcommand{\arraystretch}{1.15}
    \begin{tabular}{l l c c l}
    \hline \hline
    Region & $M$ & $\sigma_M$ & Method \\
     &  & [$10^8\,M_\odot$] & [$10^8\,M_\odot$] & References \\
    \midrule
    \multicolumn{5}{c}{NBF region (same observed area)} \\
    \midrule
    NBF & 1.85 & 0.30 & RC counts & This work\\
    NBF & 0.73 & - & Self-consistent NSD model & 1  \\
    NBF & 1.63 & - & Dynamical/model-based estimate & 2  \\
    \midrule
    \multicolumn{5}{c}{Whole NB} \\
    \midrule
    NB & 12.2 & 2.6 & Area-scaled empirical mass & This work \\
    NB & 6.9  & 2.0 & Jeans / hydrodynamical dynamical mass & 3,4  \\
    NB & 10.5 & 1.0 & Self-consistent dynamical model & 1 \\
    NB  & 14.2 & 6.0 & Fixed $M/L$ assumption & 5\\
    \hline \hline
    \end{tabular}
    \label{tab:nb_mass_comp}
    \tablebib{
    (1)~\citet{sormani+2022}; (2)~\citet{hunter+2024}; (3)~\citet{sormani+2020b}; (4)~\citet{li+2022}; (5)~\citet{launhardt+2002}.
    }
\end{table*}

\subsection{Future prospects}\label{sec:future}

As shown above, studying the stellar content of the NB is essential for obtaining a comprehensive view of the structure and processes that occur here. In this region, a joint analysis of the stellar content and gas-phase structures is not only powerful but also imperative. Also, data from new NIR facilities are essential for future studies: James Webb Space Telescope can reach down to magnitude $\sim24$ in the F210M filter (similar to Ks) with NIRCAM. This means that we could detect MS stars and measure the NB IMF. This can also allow for obtaining better estimates of the SFH of the NB, given that main-sequence turn-off stars should be at magnitude $\sim 19 -20$.

\section{Summary}\label{sec:summary}

We studied the NB stellar content to obtain new insights into its structure and to place observational constraints. We used RC stars to trace the extinction and to estimate empirically the stellar mass we can observe in this region.

We developed a novel method to estimate the extinction law towards the NB using the kinematics of the RC stars. This method assumes that the RC density peak is located at the distance of the GC, and that stars on the near side of the GC have different kinematics from those on the far side. We find a total-to-selective extinction ratio of $\mathrm{A_K/{E_{H-K}} = 1.259 \pm 0.074}$ and an absolute extinction ratio of $\mathrm{A_K/A_H = 1.794 \pm 0.046}$. These values are in the same range as previous works in the Nuclear regions of the MW, but differ from those derived using stars outside these regions. This suggests that the extinction law is dependent on the LOS. With these extinction ratios, we built a high spatial resolution reddening map, which is the most detailed for this region. Its most noticeable characteristics are the gradient on top of the Brick, which does not correlate with gas density maps, and the filamentary structures towards negative longitudes. We discussed the potential sources of errors and caveats associated with tracing structures using RC stars.

Finally, new-generation instrumentation and telescopes, such as MICADO \citep{davies+2021}, MOSAIC \cite{Hammer+2021}, and HARMONI \citep{thatte+2021} at the Extremely Large Telescope (ELT), will revolutionize the study of the NB, and it presents an interesting scientific case for future space projects, such as JASMINE \citep{kawata+2024} and GaiaNIR \citep{hobbs+2016}.

\begin{acknowledgements}
      We acknowledge the referee for their comments, which polished this article to its current version. A.~V.~N. thanks D.~Riquelme, C.~Ordenes-Huanca, B.~Acosta-Tripailao, and C.~Quezada, for their insightful discussions and helpful advice about this work. A.V.N. also acknowledges support from the National Agency for Research and Development (ANID) Scholarship Program Doctorado Nacional 2020–21201226. Also acknowledges partial funding of ANID BASAL Center for Astrophysics and Associated Technologies (CATA) FB210003, by the ANID Millennium Science Initiative, ICN12\_009 and AIM23-0001, awarded to the Millennium Institute of Astrophysics (MAS), and ESO SSDF Project 21/24. M.~Z. acknowledges support from FONDECYT Regular grant No. 1230731. E.~V. acknowledges the Excellence Cluster ORIGINS Funded by the Deutsche Forschungsgemeinschaft (DFG, German Research Foundation) under Germany’s Excellence Strategy – EXC-2094-390783311. CG acknowledges support from the Agencia Estatal de Investigaci\'on del Ministerio de Ciencia e Innovaci\'on (AEI-MCINN) under grants “At the forefront of Galactic Archaeology: evolution of the luminous and dark matter components of the Milky Way and Local Group dwarf galaxies in the {\it Gaia} era” with reference PID2023-150319NB-C21/10.13039/501100011033. F.G. gratefully acknowledges support from the French National Research Agency (ANR) funded project “MWDisc” (ANR-20-CE31-0004) and “Pristine” (ANR-18-CE31-0017). F.G. also acknowledges support from the Research fellow Gemini-ANID 2025 - 32RF250005 project, and the international Gemini Observatory, a program of NSF NOIRLab, which is managed by the Association of Universities for Research in Astronomy (AURA) under a cooperative agreement with the U.S. National Science Foundation, on behalf of the Gemini partnership of Argentina, Brazil, Canada, Chile, the Republic of Korea, and the United States of America. This work made use of a number of tools and python packages: Astropy:\footnote{http://www.astropy.org} a community-developed core Python package and an ecosystem of tools and resources for astronomy \citep{astropy-colab+2013, astropy-collab+2018, astropy-collab+2022}, Numpy \citep{harris2020array}, Scipy \citep{SciPy-NMeth2020}, the Scikit-learn library \citep{scikit-learn}, Pandas \citep{reback2020pandas}, astroML \citep{astroML}, Aladin \citep{bonnarel+2000}, and Topcat \citep{taylor2017}.
\end{acknowledgements}

\bibliographystyle{aa} 
\bibliography{NB_bibliography.bib}

\begin{appendix}
\onecolumn

\section{Observing specifications}

Here, we summarize the technical specifications of our observations, as described in \ref{sec:image_proc}. 

\begin{table*}[h!]
    \centering
    \caption{Description of the images used in this work.}
    \resizebox{\textwidth}{!}{%
    \begin{tabular}{ c c c c c c c c c c c c c c }
    \hline \hline 
    \multirow{2}{*}{\textbf{Field}} & \multirow{2}{*}{\textbf{Filter}} & \multirow{2}{*}{\textbf{Gal. Long. ($^\circ$})} & \multirow{2}{*}{\textbf{Gal. Lat. ($^\circ$)}} & \multicolumn{5}{ c}{\textbf{Short exp.}}           & \multicolumn{5}{ c }{\textbf{Long exp.}}            \\ 
    \cmidrule(lr){5-9} \cmidrule(lr){10-14} 
    & & & & \textbf{$\text{N}_{\text{obs}}$} & \textbf{$\text{N}_{\text{stack}}$}& \textbf{$\text{NDIT}$}& \textbf{$\text{DIT}$}& \textbf{$\text{t}_{\text{exp}}$} & \textbf{$\text{N}_{\text{obs}}$} & \textbf{$\text{N}_{\text{stack}}$}& \textbf{$\text{NDIT}$}& \textbf{$\text{DIT}$}& \textbf{$\text{t}_{\text{exp}}$}   \\ \hline
    \noalign{\smallskip}
    NBF054 & H    & \multirow{2}{*}{0.25841}  & \multirow{2}{*}{-0.00133} & 1      & 5 & 1 & 2~s  & 10~s      & 1    & 16  & 3 & 10~s     & 480~s     \\   
           & \Ks  &                                       & & 1      & 5 & 1 & 2~s  & 10~s      & 1    & 8   & 2 & 10~s     & 160~s     \\   
    \noalign{\smallskip}
    NBF058 & H    & \multirow{3}{*}{0.12991}  & \multirow{3}{*}{-0.00125} & 2      & 5 & 1 & 2~s  & 10~s      & 2    & 16  & 3 & 10~s     & 480~s     \\ 
           & H    &                                       & &     --          & -- & --  & --  & --    & 1    & 6   & 3 & 10~s     & 180~s     \\
           & \Ks  &                                       & & 1      & 5 & 1 & 2~s  & 10~s      & 2    & 8   & 2 & 10~s     & 160~s     \\   
    \noalign{\smallskip}
    NBF062 & H    & \multirow{2}{*}{0.00092}  & \multirow{2}{*}{-0.00119} & 1      & 5 & 1 & 2~s  & 10~s      & 2    & 16  & 3 & 10~s     & 480~s     \\   
           & \Ks  &                                       & & 1      & 5 & 1 & 2~s  & 10~s      & 2    & 8   & 2 & 10~s     & 160~s     \\   
    \noalign{\smallskip}
    NBF066 & H    & \multirow{2}{*}{-0.12833} & \multirow{2}{*}{-0.00133} & 1      & 5 & 1 & 2~s  & 10~s      & 2    & 16  & 3 & 10~s     & 480~s     \\   
           & \Ks  &                                       & & 1      & 5 & 1 & 2~s  & 10~s      & 2    & 8   & 2 & 10~s     & 160~s     \\   
    \noalign{\smallskip}
    NBF070 & H    & \multirow{2}{*}{-0.25724} & \multirow{2}{*}{-0.00151} & 1      & 5 & 1 & 2~s  & 10~s      & 1    & 16  & 3 & 10~s     & 480~s     \\   
           & \Ks  &                                       & & 1      & 5 & 1 & 2~s  & 10~s      & 1    & 8   & 2 & 10~s     & 160~s     \\   
    \noalign{\smallskip}
    \hline \hline 
    \smallskip \\
    \end{tabular}%
    }
    \label{tab:data_descript}
    
\end{table*}

\twocolumn
\section{Photometric calibration specifications} \label{sec:appendix_phot}

Here, we describe the procedure used to derive the photometric transformations between 2MASS, VVV/VVVX, and HAWK-I, accounting for the differences in filter transmission curves (color term, CT) and instrumental zero points (ZP). If the filters were identical, the color term would vanish, leaving only a zero-point offset.

The direct overlap between HAWK-I and 2MASS is limited, since the brightest unsaturated stars in HAWK-I correspond to faint stars in 2MASS. Furthermore, the different pixel sizes of HAWK-I and 2MASS (0.106 v/s 2.0 "/pix, respectively), and typical FWHM (0.25" v/s 2.5") impeded a direct cross-matching between them. To bridge this gap, we used the VVV/VVVX survey as an intermediate calibration step.

The transformation between 2MASS and VVV can be written as:

\begin{equation}\label{eq:photcal_2MASS-VVV}
    \mathrm{mag_{2MASS} = mag_{VVV} - CT\cdot(H_{VVV} - Ks_{VVV}) + ZP}
\end{equation}

\begin{figure}[]
 \includegraphics[width=\linewidth]{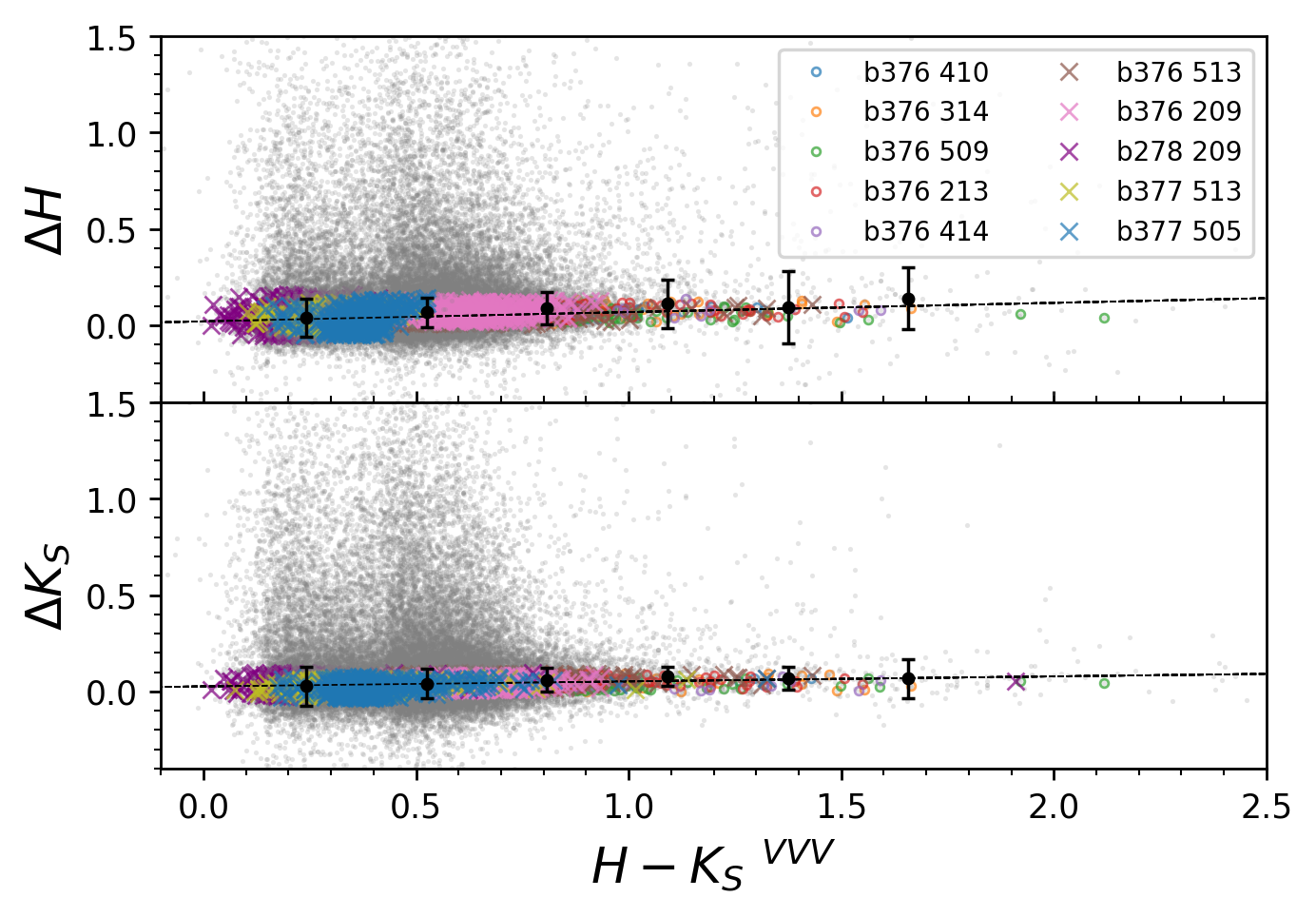}
 \caption{Color terms between 2MASS and VVV photometry. The horizontal axis shows the color in VVV system, whereas the vertical axis shows the magnitude difference VVV minus 2MASS, for H and \Ks in the top and bottom panels, respectively. The legend shows the VVV tile and chip. In gray are shown all the stars cross-matched. The colored circles represent the stars that passed the filters in photometric quality cuts (\texttt{e\_Kmag} < 0.04 and \texttt{e\_Hmag} < 0.04) and the sigma-clipping procedure. Each color represents a different VVV chip. The black dots represent the median within each bin, and the error bar is the median absolute deviation within that bin. The final number of stars used to find the color terms is 8\,949. The color term for H is $CT_H = 0.048$, whereas for \Ks\ is $CT_{Ks} = 0.026$}
 \label{fig:CT_2MASS_VVV}
\end{figure}

To find the CT on Eq.~\ref{eq:photcal_2MASS-VVV}, we cross-matched VVV/VVVx with 2MASS around coordinates $\mathrm{(l,b)=(2.0^\circ, 4.0^\circ)}$, where the crowding is low but the differential extinction is large enough to provide a non-negligible color baseline, as shown in Fig.~\ref{fig:CT_2MASS_VVV}. Here, we selected stars from 9 VVV chips from tiles b376, and b377. We also added one chip from we tile b278, where Baade's Window is located in VVV data\citep[the map and positions of VVV/VVVX tiles can be found on Fig A.1 and Table A.1 of][respectively]{saito+2024}. We found 46\,471 stars in common with 2MASS. We chose these chips in tiles b376 and b377 because they are contiguous detectors with a color range in $\mathrm{H-K}$ similar to the one in the HAWK-I data. We also included one chip from tile b278 because it covers part of Baade's Window and adds statistics at the low-reddening end, as seen in Fig. \ref{fig:CT_2MASS_VVV}. We computed the CT by separating the VVV $\mathrm{H-K_s}$ color in equally spaced bins, masking the data with errors $\mathrm{\texttt{e\_Kmag}} < 0.04$ and $\mathrm{\texttt{e\_Hmag}} < 0.04$. For each bin, we computed the median of the magnitude difference between VVV and 2MASS, and assigned the median absolute deviation as the bin uncertainty. Next, we iterated the sigma-clipping procedure 3 times with a threshold of $0.8\sigma$ to minimize the influence of outliers. Finally, we fitted the data with a first-degree polynomial using least-squares minimization. We interpreted the slope as the CT. With these constraints, we kept 8\,949 stars to calibrate the data: 4\,721 for H, and 4\,208 for \Ks\.  The CT are $\mathrm{CT_H = 0.048 \pm 0.004}$ and $\mathrm{CT_{Ks} = 0.026 \pm 0.007}$. To estimate the uncertainties in the CT, we resampled the data 1\,000 times and repeated the procedure described previously. The reported uncertainty corresponds to $2\sigma$ of the CTs distribution.

Next, we computed the ZPs in the overlapping region of NBFs. We cross-matched the chips b334\_203, b333\_515, b333\_215, and b333\_511 with 2MASS. We found a total of 32\,832 stars in common in this area. We applied the quality cuts over $\mathrm{\texttt{e\_Kmag}} < 0.1$ and $\mathrm{\texttt{e\_Hmag}} < 0.1$. Also, we ensured a good overlap by imposing that $10 < \mathrm{\texttt{Kmag}} < 12$ and $12 < \mathrm{\texttt{Hmag}} < 14$. With these constraints, we kept 8\,675 stars. We use a higher error threshold because, if we set it to $0.04$ as before, the number of 2MASS stars falls to $\sim$~1\,000. We computed the ZP in H and \Ks\ using the mode of the magnitude difference distribution between VVV/VVVX and 2MASS. The ZPs are found in Table~\ref{tab:ZP_2MASS_VVV}. Finally, we applied Eq.~\ref{eq:photcal_2MASS-VVV} to the VVV data in the NBF region to calibrate the VVV to 2MASS. 

\begin{table}
    \centering
    \caption{Zero-points between 2MASS and VVV overlapping with NBFs.}
    \begin{tabular}{lcc}
    \hline
    Field & $ZP_{K_s}$ & $ZP_H$ \\
    \hline
    zb334\_203 & $-0.025 \pm 0.014$ & $0.058 \pm 0.026$ \\
    zb333\_215 & $-0.034 \pm 0.014$ & $0.150 \pm 0.025$ \\
    zb333\_515 & $-0.040 \pm 0.012$ & $0.146 \pm 0.022$ \\
    zb333\_511 & $-0.021 \pm 0.012$ & $0.103 \pm 0.021$ \\
    \hline
    \end{tabular}
    \label{tab:ZP_2MASS_VVV}
\end{table}

We calibrated the NBFs photometry using the 2MASS-calibrated VVV catalog as a reference, by fitting the equation:

\begin{equation}\label{eq:photcal_VVV-NBF}
    \mathrm{mag_{VVV} = mag_{NBF} - CT^{NBF}\cdot(H_{NBF} - Ks_{NBF}) + ZP}
\end{equation}

We selected common stars with 2MASS magnitude $12 < \mathrm{K_S} < 16$, that are not saturated in NBF but well detected in VVV, and independently calibrated each NBF chip. The procedure was similar to that with 2MASS: we binned the data by color, applied an iterative 1-sigma clipping, and fitted a first-degree polynomial to the median of each bin using least-squares minimization. The uncertainties were estimated by running 1\,000 bootstrapping iterations, and considering 2-sigma of the resampling distribution. The slope of the fit is interpreted as the CT, whereas the free term is the ZP. If the value of the CT is consistent with zero within the uncertainties, we did not apply the CT and only applied the ZP, which is true for the CTs in K, as shown in Table~\ref{tab:NBF_zp_ct_grouped_centered}. After applying Eq.~\ref{eq:photcal_VVV-NBF} to each HAWK-I detector, we obtained the photometry shown in Fig.~\ref{fig:CMD_NPL}.

\begin{table*}[ht]
\caption{Photometric ZP and CT for NBF fields.}
\centering
\label{tab:NBF_zp_ct_grouped_centered}
\begin{tabular}{cc cc cc}
\hline \hline
\multirow{2}{*}{NBF field} &
\multirow{2}{*}{chip} &
\multicolumn{2}{c}{H filter} &
\multicolumn{2}{c}{K$_\mathrm{S}$ filter} \\
\cmidrule(lr){3-4} \cmidrule(lr){5-6}
 &  & ZP & CT & ZP & CT \\
\midrule
NBF054 & Q1 & $3.73 \pm 0.01$ & $-0.083 \pm 0.010$ & $2.97 \pm 0.02$ & $-0.010 \pm 0.012$ \\
       & Q2 & $3.76 \pm 0.01$ & $-0.084 \pm 0.016$ & $3.00 \pm 0.01$ & $-0.020 \pm 0.012$ \\
       & Q3 & $3.77 \pm 0.01$ & $-0.084 \pm 0.015$ & $3.01 \pm 0.01$ & $-0.012 \pm 0.015$ \\
       & Q4 & $3.77 \pm 0.01$ & $-0.079 \pm 0.028$ & $2.99 \pm 0.01$ & $-0.018 \pm 0.015$ \\
\midrule
NBF058 & Q1 & $3.72 \pm 0.01$ & $-0.084 \pm 0.022$ & $2.88 \pm 0.02$ & $-0.019 \pm 0.024$ \\
       & Q2 & $3.69 \pm 0.02$ & $-0.104 \pm 0.038$ & $2.90 \pm 0.03$ & $-0.014 \pm 0.026$ \\
       & Q3 & $3.76 \pm 0.01$ & $-0.100 \pm 0.018$ & $3.02 \pm 0.02$ & $-0.021 \pm 0.013$ \\
       & Q4 & $3.70 \pm 0.01$ & $-0.078 \pm 0.014$ & $2.96 \pm 0.02$ & $-0.005 \pm 0.011$ \\
\midrule
NBF062 & Q1 & $3.76 \pm 0.01$ & $-0.077 \pm 0.026$ & $2.97 \pm 0.03$ & $-0.017 \pm 0.028$ \\
       & Q2 & $3.74 \pm 0.03$ & $-0.128 \pm 0.031$ & $2.95 \pm 0.04$ & $-0.034 \pm 0.032$ \\
       & Q3 & $3.76 \pm 0.03$ & $-0.081 \pm 0.019$ & $3.06 \pm 0.03$ & $-0.027 \pm 0.017$ \\
       & Q4 & $3.81 \pm 0.02$ & $-0.096 \pm 0.014$ & $3.03 \pm 0.02$ & $-0.008 \pm 0.016$ \\
\midrule
NBF066 & Q1 & $3.74 \pm 0.02$ & $-0.093 \pm 0.022$ & $2.91 \pm 0.04$ & $+0.006 \pm 0.025$ \\
       & Q2 & $3.80 \pm 0.03$ & $-0.097 \pm 0.048$ & $2.94 \pm 0.02$ & $-0.015 \pm 0.038$ \\
       & Q3 & $3.79 \pm 0.02$ & $-0.052 \pm 0.034$ & $3.04 \pm 0.02$ & $0.000 \pm 0.024$ \\
       & Q4 & $3.76 \pm 0.02$ & $-0.095 \pm 0.016$ & $3.04 \pm 0.02$ & $-0.026 \pm 0.012$ \\
\midrule
NBF070 & Q1 & $3.75 \pm 0.03$ & $-0.074 \pm 0.043$ & $2.93 \pm 0.02$ & $-0.028 \pm 0.026$ \\
       & Q2 & $3.79 \pm 0.01$ & $-0.075 \pm 0.033$ & $2.97 \pm 0.02$ & $-0.004 \pm 0.021$ \\
       & Q3 & $3.83 \pm 0.01$ & $-0.084 \pm 0.022$ & $3.01 \pm 0.03$ & $-0.007 \pm 0.017$ \\
       & Q4 & $3.80 \pm 0.02$ & $-0.079 \pm 0.024$ & $2.98 \pm 0.02$ & $-0.013 \pm 0.024$ \\
\hline \hline
\end{tabular}\\
\tablefoot{Each column displays the field name, the chip number corresponding to the HAWK-I detector number, the ZP and CT for the H filter, and the ZP and CT for the K$_\mathrm{S}$ filter, along with their associated uncertainties, which are 2 standard deviations of the resampled distribution of parameters.}
\label{tab:ZP_CT}
\end{table*}

\section{Reddening map (Table at CDS)}
\label{app:reddening_map}

Table~\ref{tab:redd_map} lists the selective extinction $E(H-K_\mathrm{s})$ across the Nuclear Bulge, derived from kinematically-selected red clump stars observed with HAWK-I/VLT. Column~1 gives the Galactic longitude, Column~2 the Galactic latitude, Column~3 the selective extinction $E(H-K_\mathrm{s})$, and Column~4 an interpolation flag indicating whether the value was directly measured (0) or interpolated from neighbouring bins (1). The full table is available at the CDS.

\begin{table}
\caption{Reddening map of the Nuclear Bulge (excerpt).}
\label{tab:redd_map}
\centering
\begin{tabular}{cccc}
\hline\hline
$l$ & $b$ & $E(H-K_\mathrm{s})$ & Flag \\
(deg) & (deg) & (mag) & \\
\hline
$-$0.3248 & $-$0.0686 & 1.647 & 1 \\
$-$0.3248 & $-$0.0672 & 1.710 & 0 \\
$-$0.3248 & $-$0.0658 & 1.407 & 0 \\
$-$0.3248 & $-$0.0644 & 1.559 & 0 \\
$-$0.3248 & $-$0.0631 & 1.631 & 0 \\
\vdots & \vdots & \vdots & \vdots \\
\hline
\end{tabular}
\tablefoot{This table is available in its entirety (50\,000 rows) 
in machine-readable form at the CDS.}
\end{table}

\end{appendix}

\end{document}